\documentclass[12pt,preprint,numberedappendix,twocolappendix]{emulateapj}

\newcommand\bibinc{n}		

\usepackage{subeqnarray}
\usepackage{amsmath}
\usepackage{hyperref}
\usepackage{ulem}
\usepackage{stmaryrd}

\hypersetup{colorlinks=true,linkcolor=black,citecolor=blue,urlcolor=black}

\usepackage{natbib}
\def\tt{\bf   }

\newcommand{\Eq}[1]{Equation\,(\ref{#1})}

\newcommand{\Sec}[1]{Section~\ref{#1}}

\newcommand{\Fig}[1]{Figure~\ref{#1}}

\def \Kzz {K_{\mathrm{zz}}}

\newcommand {\twostr} {{\ttfamily TWOSTR}}
\newcommand {\disort} {{\ttfamily DISORT}}
\newcommand {\mitgcm} {{\ttfamily MITgcm}}

\setcitestyle{citesep={,}}
\begin{document}

\slugcomment{Accepted at ApJ}

\shorttitle{Vertical Mixing in Hot Jupiter Atmospheres}
\shortauthors{Komacek, Showman, \& Parmentier}

\title{Vertical Tracer Mixing in Hot Jupiter Atmospheres}
\author{Thaddeus D. Komacek$^{1}$, Adam P. Showman$^{2,3}$, and Vivien Parmentier$^{4}$} \affil{$^1$Department of the Geophysical Sciences, The University of Chicago, Chicago, IL, 60637 \\ 
 $^2$Department of Atmospheric and Oceanic Sciences, School of Physics, Peking University, Beijing, 100871, China \\
 $^3$Lunar and Planetary Laboratory and Department of Planetary Sciences,
 University of Arizona, Tucson, AZ, 85721 \\
$^4$Department of Physics, Oxford University, OX1 2JD, United Kingdom \\
\url{tkomacek@uchicago.edu}} 
\begin{abstract}
Aerosols appear to be ubiquitous in close-in gas giant atmospheres, and disequilibrium chemistry likely impacts the emergent spectra of these planets. Lofted aerosols and disequilibrium chemistry are caused by vigorous vertical transport in these heavily irradiated atmospheres. Here we numerically and analytically investigate how vertical transport should change over the parameter space of spin-synchronized gas giants. In order to understand how tracer transport depends on planetary parameters, we develop an analytic theory to predict vertical velocities and mixing rates ($\Kzz$) and compare the results to our numerical experiments. We find that both our theory and numerical simulations predict that, if the vertical mixing rate is described by an eddy diffusivity, then this eddy diffusivity $\Kzz$ should increase with increasing equilibrium temperature, decreasing frictional drag strength, and increasing chemical loss timescales. We find that the transition in our numerical simulations between circulation dominated by a superrotating jet and that with solely day-to-night flow causes a marked change in the vertical velocity structure and tracer distribution. The mixing ratio of passive tracers is greatest for intermediate drag strengths that corresponds to this transition between a superrotating jet with columnar vertical velocity structure and day-to-night flow with upwelling on the dayside and downwelling on the nightside. Lastly, we present analytic solutions for $\Kzz$ as a function of planetary effective temperature, chemical loss timescales, and other parameters, for use as input to one-dimensional chemistry models of spin-synchronized gas giant atmospheres.
\end{abstract}
\section{Introduction}
\label{sec:intro}
\subsection{Aerosols appear to be ubiquitous in hot Jupiter atmospheres}
Close-in extrasolar giant planets (colloquially known as ``hot Jupiters'') are intensely irradiated, with equilibrium temperatures in excess of $1000 \ \mathrm{K}$. Even though the atmospheres of these planets are extremely hot, it is clear from transit observations that their spectra are affected by extra opacity sources beyond pure molecular absorption. This was hinted at by the first atmospheric detection of a hot Jupiter via sodium absorption in visible light observations of HD 209458b \citep{Charbonneau:2002}. \citeauthor{Charbonneau:2002} found that the transit depth was not as large as expected from the clear atmosphere models of \cite{Seager2000}. One explanation for this lowered transit depth is a cloud deck forming at high altitudes in the atmosphere of the planet and reducing the amplitude of absorption features. A decade later, it was found through \textit{Hubble} and \textit{Spitzer} observations that HD 189733b also does not show evidence for molecular features from the ultraviolet to infrared, pointing toward the presence of aerosols in its atmosphere \citep{desert2011,gibson2011,pont2013}. In the absence of atmospheric dynamics and the resulting vertical transport, aerosols would settle to layers below the photosphere. As a result, knowledge of the large-scale vertical transport potentially constrains the presence of observable aerosols in hot Jupiter atmospheres.  \\ 
\indent The state of the art of transit observations of hot Jupiters includes a large sample of planets with both near-infrared \textit{HST-WFC3} spectrophotometry and \textit{Spitzer} mid-infrared photometry (summarized in \citealp{Sing:2015a}). \citeauthor{Sing:2015a} showed that the measured amplitude of water-vapor absorption features from \textit{WFC3} correlated with the altitude difference between the near-to-mid IR continuum in a manner that was best matched by models with either a Rayleigh-scattering haze opacity or a grey cloud\footnote{From here onward we use ``cloud'' to refer to an equilibrium condensate and ``haze'' to refer to a solid or liquid photochemical or disequilibrium chemical product.} opacity.
Analyzing the results of \cite{Sing:2015a} and complementary data, \cite{Heng:2016} and \cite{Stevenson:2016} found that the cloudiness of this sample of planets may decrease with increasing incident stellar flux. This potential observational trend of decreasing cloudiness with increasing temperature motivates further analysis of how the cloud distribution of hot Jupiters changes with planetary parameters.   \\
\indent Though infrared observations in transmission show strong evidence for the presence of clouds in hot Jupiter atmospheres, they only give insight on the cloud distribution at the terminator of the planet. Beginning with observations of Kepler-7b \citep{Demory_2013}, \textit{Kepler} visible-light phase curves have enabled inferences of the distribution of clouds across longitude. To date, there are eleven hot Jupiters with visible light phase curve observations \citep{Angerhausen2015,Esteves:2015,Shporer:2015a}. These observations have shown that the offset of the maximum of the phase curve correlates with incident stellar flux \citep{Esteves:2015,Parmentier:2015} such that the cooler planets have westward offsets and the two hottest planets in the sample have eastward offsets. The theoretical work of \cite{Showman_Polvani_2011} predicts that hot Jupiters should have a superrotating (eastward) jet at the equator. This eastward jet results in eastward offsets of the infrared hot spot, which have been predicted by numerical simulations \citep{showman_2002,Showman:2008} and found in observed phase curves \citep{Knutson_2007,Knutson:2009,Knutson:2009a,Cowan:2012,Knutson:2012,Stevenson:2014,Zellem:2014,Wong:2015,Wong:2015a}. \\
\indent The reason why that many \textit{Kepler} observations show westward phase curve offsets is explained by the models of \cite{Hu:2015} and \cite{Parmentier:2015}. The nightside of hot Jupiters are cold, which promotes cloud formation. As hot Jupiters have a superrotating equatorial jet, winds transport cold and cloudy air from the nightside towards the region of the dayside west of the substellar point. Thus, clouds form on the dayside west of the substellar point. As these winds cross eastward over the dayside while absorbing starlight, the temperature rises -- with the peak temperatures typically occurring east of the substellar point -- and the cloud sublimates. Hence, the regions east of the substellar point are more likely to be cloud free. In summary, one would expect more cloud formation westward of the substellar point, with a corresponding greater albedo in the visible westward of the substellar point. Conversely, thermal emission is more important eastward of the substellar point, and if the planet becomes hot enough this can start to dominate over the cloud albedo affect in the visible \citep{Hu:2015,Parmentier:2015}. This basic scenario explains the transition in the sign of the visible-wavelength offset with increasing planetary effective temperature. \\
\indent Most recently, the visible phase curve offset of HAT-P-7b has been found to be strongly time-variable \citep{Mooij2016}, flipping from east to west over a timescale of tens to hundreds of days. There is yet no detailed explanation for what causes these extreme variations in the phase curve of HAT-P-7b, but it is likely tied to clouds \citep{Mooij2016}, magnetohydrodynamic effects \citep{Rogers:2017}, or a combination of the two. \\
\indent Secondary eclipse observations of hot Jupiters measure the flux from only the dayside of the planet and hence provide a constraint on the planetary albedo. If the observation is taken at visible wavelengths where the planetary emission is small relative to the incident stellar flux these observations constrain the geometric albedo, giving a measure of how reflective the planet is. A few notable hot Jupiters have been found to exhibit high geometric albedos, including HD 189733b \citep{evans2013} and Kepler-7b \citep{Demory_2013}. The high geometric albedos observed for many hot Jupiters in the optical (reaching $0.43$ for HD 209458b, $0.37$ for HD 189733b, and $0.34$ for Kepler-7b at wavelengths $\le 0.8 \ \mu\mathrm{m}$, \citealp{Schwartz:2015})  points toward the presence of aerosols on their daysides. However, the bulk of hot Jupiters observed have low geometric albedos, in many cases lower than the bond albedos of planets with similar equilibrium temperatures \citep{esteves2013,Schwartz:2015}. This may be due to cold trapping of silicates for planets with equilibrium temperatures $\lesssim 1600~\mathrm{K}$ preventing vertical lofting of silicate aerosols \citep{Parmentier:2015} or an optical absorber reducing the albedo \citep{Schwartz:2015}.  \\
\indent Though there is convincing evidence that aerosols are present in most observed atmospheres of close-in extrasolar giant planets, a detailed understanding of these particles (e.g. their chemical composition, particle size, and three-dimensional distribution) is lacking. Based on the modeled temperature-pressure profiles of hot Jupiters, it is expected that the key potential condensates are (magnesium) silicates, iron and iron oxides, calcium-aluminum bearing minerals (e.g. CaTiO$_3$), and magnesium sulfides and oxides \citep{Sing:2015a,Helling:2016,Parmentier:2015,Wakeford:2017}. These cloud species 
would easily settle out of the atmosphere were there no vertical transport keeping them aloft \citep{Ackerman2001,parmentier_2013,Gao:2018}. As a result, determining the strength of vertical transport of aerosol particles in hot Jupiter atmospheres is a key component for any framework to understand cloud formation in their atmospheres. \\
\indent The current state of the art for cloud modeling of exoplanets is to couple a cloud model to a dynamical general circulation model (GCM). As a first step, \cite{parmentier_2013} coupled passive aerosol tracers which mimicked clouds that did not radiatively interact with the circulation. The radiative effect of clouds has been modeled in two ways, with increasing complexity. Both \cite{Lee:2015} and \cite{Parmentier:2015} post-processed a dynamical model with a model to calculate the cloud composition and how clouds affect observed properties of the planet. However, they utilized different GCMs, as \cite{Lee:2015} used the fully compressible Navier-Stokes model of \cite{Dobbs-Dixon:2013} with a simplified form of line-by-line radiative transfer while \cite{Parmentier:2015} used the primitive equation model of \cite{Showmanetal_2009} (see also \citealp{Lewis:2010,Kataria2016}) with a more realistic correlated-k radiative transfer scheme. Additionally, they utilized different cloud models, where \cite{Lee:2015} applied the microphysical cloud model of \cite{Woitke2003} and \cite{Woitke2004}, and \cite{Parmentier:2015} used an equilibrium cloud scheme assuming that the particle radius is constant and the mixing ratio of the cloud material is constant with altitude. Notably, because their dynamical models covered a wide range of incident stellar fluxes, \cite{Parmentier:2015} determined that the cloud composition must change with equilibrium temperature, hinting toward the presence of a deep cold-trap. \\
\indent Improving upon the model of \cite{Lee:2015}, \cite{Lee:2016} and \cite{Lines:2018} directly coupled the radiative effects from cloud tracer particles with the dynamical cores of \cite{Dobbs-Dixon:2013} and \cite{Mayne:2014}, respectively. Additionally, note that such coupling has also been done for super-Earths by \cite{Charnay:2015} using an equilibrium cloud framework. In this work, we study passive (i.e., non-radiatively interacting) aerosol tracers with a range of fixed particle radii. We do so because simplifying the microphysics is crucial for obtaining a deep physical, mechanism-based, and analytical understanding of the transport process. This also allows us to cover a wide parameter space, studying the transport of aerosol particles of a wide range of sizes (from $0.01-10 \ \mu\mathrm{m}$) in atmospheres covering a three orders of magnitude range in incident stellar flux and more than four orders of magnitude in large-scale drag.
\subsection{Further impacts of vertical mixing: disequilibrium chemistry}
\indent In the absence of dynamics, the atmospheric chemistry is expected to be determined solely by its pressure, temperature, and elemental abundances. However, in the presence of dynamics, the interaction of dynamics and chemistry can lead to additional disequilibrium effects. That is, dynamics can mix chemical species faster than they can react toward equilibrium, resulting in species that are in disequilibrium. Understanding the nature of chemical disequilibrium hence requires an understanding of tracer transport.  \\
\indent Several methods have been adopted to study this interaction between dynamics and chemistry. Most prominent are one-dimensional chemical models that solve a vertical diffusion equation, where the dynamical mixing that would occur in a real atmosphere is parameterized as a diffusive process. Given this simplified dynamical framework, these models self-consistently solve for the abundances of given chemical species using a thermo/photochemical model \citep{Moses:2011,Visscher:2011,Venot:2012,Moses2013}. These studies have shown that disequilibrium chemistry has an impact on the chemical abundances and hence emergent spectra of hot Jupiters. \\
\indent Disequilibrium effects have also been modeled in GCMs by coupling simplified chemical tracers that relax toward chemical equilibrium over a chemical relaxation timescale \citep{Cooper:2006,Drummond:2018aa,Drummond:2018ab,Mendonca:2018aa,Zhang:2018aa}, or by prescribing species in chemical disequilibrium \citep{Steinrueck:2018aa}. These GCM studies have found that horizontal transport is also important alongside vertical transport for determining the abundance of disequilibrium species. Notably, the strong horizontal transport in hot Jupiter atmospheres can cause molecules such as CO and H$_2$O to have near-uniform abundances along the equatorial regions \citep{Cooper:2006,Mendonca:2018aa}. This horizontal transport can strongly affect the emergent spectrum of hot Jupiters as a function of orbital phase \citep{Mendonca:2018aa,Steinrueck:2018aa}. Additionally, disequilibrium effects can be important to determine atmospheric chemical abundances of brown dwarfs and directly-imaged giant planets \citep{Line:2013,Bordwell:2018}. \\
\indent Even simpler models without any explicit representation of mixing often simply use a ``quench approximation'' in the vertical direction. The quench appproximation assumes that the atmosphere is in chemical equilibrium where the chemical reaction timescales are much shorter than the dynamical timescale (i.e., the vertical mixing timescale), and the chemical abundances are in disequilibrium where the dynamical timescale is shorter than the chemical inter-conversion timescale \citep{Smith:1998}. This simple timescale comparison allows one to approximately identify the quench level, below which the abundances are close to equilibrium and above which the species are quenched (or at least strongly affected by dynamics). These one-dimensional models lack dynamics and so parameterize the vertical mixing rate that sets the quenching level as a one-dimensional diffusion coefficient, which is characterized by a vertical eddy diffusivity $\Kzz$. This implies a diffusion (or dynamical) timescale $\tau_{\mathrm{dyn}} \sim H^2/\Kzz$, where $H$ is the atmospheric scale height\footnote{See \citealp{Bordwell:2018} for a more rigorous treatment of the relevant chemical length scale.}. In this work, we develop a theoretical prediction for $\Kzz$ that can be utilized in both quench approximation and thermo/photochemical kinetics models. 

\subsection{The utility of an analytic theory for vertical mixing rates}
It is clear from the above discussion that understanding vertical transport is necessary to predict the impact of aerosols and disequilibrium chemistry on observations of hot gas giants. However, there is a lack of predictive theory for the vertical mixing rate and how it should vary with properties of the planet, e.g. incident stellar flux, rotation rate, surface gravity, and composition. Though estimates of $\Kzz$ from mixing length theory have been used as input for cloud models \citep{Ackerman2001,Gao:2018}, these estimates only directly apply in the convective interior of the planet. As hot Jupiters have deep radiative (stably stratified) envelopes (due to the strong stellar irradiation of their atmospheres), estimates of $\Kzz$ from mixing-length theory do not apply to the atmospheres of hot Jupiters. Instead, a dynamical model developed from first principles is necessary to estimate global-scale vertical velocities and hence mixing rates, while an understanding of waves and small-scale phenomena is required to understand local mixing \citep{Fromang:2016,Menou:2018aa}. Many sophisticated GCMs have been applied to study the circulation of hot Jupiter atmospheres \citep{showman_2002,Cooper:2005,Menou:2009,Showmanetal_2009,Showman_2009,Rauscher:2010,Thrastarson:2010,Heng:2011,Showman_Polvani_2011,perna_2012,Rauscher_2012,Dobbs-Dixon:2013,Mayne:2014,Kataria2016,Komacek:2017,Drummond:2018ab,Mendonca:2018}, but as yet there are only a handful that include tracers to understand the mixing of chemical and/or cloud species \citep{Cooper:2006,parmentier_2013,Lee:2016,Drummond:2018ab,Mendonca:2018aa,Zhang:2018aa}. \\
\indent In a series of two papers, \cite{Zhang:2017,Zhang:2018aa} provided an analytic theory of vertical transport in the atmospheres of tidally-locked planets and explored how vertical transport should behave over a wide parameter space. \citeauthor{Zhang:2017} developed an analytic theory for the vertical transport of tracers that were assumed to react chemically following an idealized relaxation scheme, which attempts to relax the chemical abundance toward a prescribed, spatially varying chemical-equilibrium abundance.
They derived theory for the separate cases of tracers whose chemical equilibrium abundances are constant horizontally but vary with height, and chemical tracers whose equilibrium abundance is large on the dayside and small on the nightside (representing photochemical production on the daysides of hot Jupiters, for example).  
\citeauthor{Zhang:2017} found that the analytic theory provided a good match to a large suite of circulation models with passive tracers, including both simplified two-dimensional models of planets with a prescribed mass streamfunction and three-dimensional GCMs of tidally-locked planets where the day-night thermal forcing is driven by an idealized Newtonian heating/cooling scheme. \\
\indent The actual three-dimensional dynamics of hot Jupiters is not diffusive, instead transporting chemical tracers vertically through vertical advection associated with a variety of different dynamical processes, including large-scale overturning circulations and wave breaking. Even though the circulation is not diffusive, it is practically useful to parameterize the mixing rate using a vertical diffusivity $\Kzz$, which is the diffusivity that would be needed for a one-dimensional diffusive model with the same vertical profile of horizontal-mean tracer abundance to transport the same tracer flux as a three-dimensional model transports through dynamical motions. This $\Kzz$ is valid for tracers that are mixed from the bottom upward, for instance clouds that condense deep in the atmosphere and are transported to higher altitudes. \cite{Zhang:2018aa} utilized the predictions for the wind speeds of tidally-locked gas giants from \cite{Komacek:2015} and \cite{Zhang:2016} to predict such effective diffusivities ($K_\mathrm{zz}$) from first principles. \cite{Zhang:2018aa} found reasonably good agreement between this first-principles analytic theory and the actual dynamical mixing rates diagnosed from a suite of GCM experiments. \\
\indent Here we set out to build upon \cite{Zhang:2017,Zhang:2018aa} to determine from first principles how vertical transport in the atmospheres of spin-synchronized gas giants should scale with relevant parameters, including their incident stellar flux, atmospheric pressure level, rotation rate, and potential strength of macroscopic frictional drag. Our model uses the same conceptual framework as \cite{Zhang:2017,Zhang:2018aa}, but we improve upon their work in a variety of ways. First, our model is more sophisticated, as we use double-grey radiative transfer (and include some models with more complex correlated-k radiative transfer) rather than an idealized Newtonian heating/cooling scheme to drive the day-night temperature differences (and the resulting circulation). Second, we not only explore the behavior of a simple chemical relaxation scheme, where the chemical tracers relax toward a prescribed chemical equilibrium abundance over time, but also explore a scheme where the tracer represents particles that settle vertically. Finally, we explore the behavior over a much wider range of planetary parameters that affect the circulation, exploring the effects of varying incident stellar flux and frictional drag strength over many orders of magnitude. This macroscopic frictional drag may be due to Lorentz forces \citep{Perna_2010_1,batygin_2013,Rauscher_2013,Rogers:2020,Rogers:2014} or shear instabilities and macroscopic turbulence \citep{Li:2010,Youdin_2010,Fromang:2016}. Note that double-diffusive shear instabilities can enhance vertical transport in hot Jupiter atmospheres \citep{Menou:2018aa}, but we do not incorporate this effect in our simulations. Additionally, in this work we apply the theory of \cite{Komacek:2015} and its extension by \cite{Zhang:2016} to understand our results. This theory has been shown to match well the characteristic horizontal and vertical wind speeds from GCMs. We then relate these wind speeds to vertical mixing rates using a similar approach to \cite{Zhang:2018aa}. \\
\indent In this paper, we explore vertical transport in hot Jupiter atmospheres by coupling passive tracers (subject to appropriate sources/sinks) to dynamics in order to understand in detail how vertical transport varies with incident stellar flux, drag strength, particle size, and chemical interconversion timescales. We vary the incident stellar flux and drag strength because these parameters have the largest effect on the global circulation, as weak irradiation and strong drag both greatly damp atmospheric winds \citep{Komacek:2017,Koll:2017}. However, note that we use a simplified Rayleigh drag parameterization that crudely represents magnetic effects and/or large-scale instabilities which lead to turbulence. We do not vary the planetary surface gravity in this work. However, note that aerosol tracer transport would be greatly affected by varying gravity, as the terminal velocity of aerosol particles scales directly with gravity. As a result, vertical transport of aerosol tracers is expected to be weaker on higher-gravity planets.  \\
\indent We explore vertical transport in both analytic theory and numerical GCM experiments, and compare the two, showing that the derived trends for the vertical mixing rate with varying incident stellar flux and drag strength are similar. Specifically, we use the method of \cite{parmentier_2013} to calculate the effective globally-averaged $\Kzz$ from these simulations, finding as in \cite{Zhang:2017} that $\Kzz$ does not have a monotonic dependence on the chemical timescale. For our analytic theory, we apply the model of \cite{Holton:1986}, developed for mixing in Earth's stratosphere, to estimate $\Kzz$ from our predicted vertical wind speeds. We find good agreement between the $\Kzz$ calculated from our GCM experiments and our theoretical predictions. As a result, this prediction for $\Kzz$ can be used as input for future cloud and chemistry models of hot Jupiter atmospheres. \\
\indent This paper is organized as follows. In \Sec{sec:numerics}, we present results from our double-grey GCM showing how the calculated vertical velocities, tracer mixing ratios, and the $\Kzz$ we calculate from our GCM vary with the incident stellar flux, drag strength, particle size, and chemical timescale. 
In \Sec{sec:theory} we describe our theory for the vertical mixing rate and compare our predictions for vertical velocities and $\Kzz$ with those calculated from our suite of GCM experiments. In \Sec{sec:disc} we discuss potential applications of our theory to one-dimensional models of spin-synchronized gas giant atmospheres along with the limitations of such an application. \Sec{sec:cook} describes how to use our analytic theory to estimate vertical mixing rates. Lastly, we delineate our conclusions in \Sec{sec:conclusions}.
\section{Double-grey general circulation models with passive tracers}
\label{sec:numerics}
\subsection{Model setup}
\label{sec:setup}
\subsubsection{Dynamics and radiative transfer}
The GCM experiments in this work use the same basic numerical setup (dynamical core, radiative transfer, and planet properties) as in \cite{Komacek:2017}. This includes solving the equations of fluid motion using the \mitgcm \ \citep{Adcroft:2004} as a dynamical core, along with double-grey radiative transfer adapted from the \twostr \ mode of \disort \ \citep{Stamnes:2027,Kylling:1995}. For the dynamics, we solve the hydrostatic primitive equations (Equations 1-5 in \citealp{Komacek:2015}) in three dimensions. We incorporate a Rayleigh drag using the setup of \cite{Komacek:2015}, where the drag strength (characterized in our simulations by a drag timescale, which is the inverse of the drag strength) in the free atmosphere is set to be constant with height. We also incorporate the basal drag scheme developed by \cite{Liu:2013}, which couples the atmosphere to the deeper interior and helps to ensure that the results of our simulations are insensitive to initial conditions. The basal drag scheme is set at pressures greater than 10 bars and has a timescale of 10 days at the bottom of the domain, increasing to infinity (corresponding to no drag) at pressures less than 10 bars (see Figure 2 of \citealp{Komacek:2015}). As in \cite{Komacek:2017}, we vary the spatially constant drag in the atmosphere by varying the characteristic drag timescale in the atmosphere $\tau_\mathrm{drag}$ from $10^3 \mathrm{s} - \infty$ in order-of magnitude intervals (up to $10^7 \ \mathrm{s}$, and including $\infty$, corresponding to no drag). \\
\indent The radiative forcing in our GCM experiments considers a planet in synchronous rotation, with a permanent irradiated dayside and a nightside in permanent darkness. This instellation pattern causes strong heating variations from the dayside to the nightside, which drives atmospheric circulation. We use a double-grey radiative transfer scheme that uses the plane-parallel two stream approximation of radiative transfer with the hemispheric closure, which ensures energy conservation \citep{Pierrehumbert:2010}. We utilize only two grey absorption coefficients, one in the visible and one in the infrared, with the same values as in \cite{Komacek:2017}. The visible absorption coefficient is set to a constant ($\kappa_v = 4 \times 10^{-4} \ \mathrm{m}^2 \ \mathrm{kg}^{-1}$), while the infrared absorption coefficient is taken to be power-law in pressure ($\kappa_{th} = 2.28 \times 10^{-6} \ \mathrm{m}^2 \ \mathrm{kg}^{-1} (p/1 \mathrm{Pa})^{0.53}$) in order to mimic the effects of collision-induced absorption \citep{Arras:2006kl,Heng:2011a,Rauscher_2012}, and scattering is neglected. The opacities are chosen to match the analytic solutions of \cite{Parmentier:2014,Parmentier:2014a} for HD 209458b. We use the same opacity setup across the suite of simulations, in order to isolate the effect of varying planetary parameters on mixing. As a result, these simulations are not meant to be representations of individual hot Jupiters. As in \cite{Komacek:2017}, we vary the incident stellar flux at the substellar point such that the equilibrium temperature $T_\mathrm{eq} = \left[F_{\star}/(4\sigma)\right]^{1/4}$, where $F_\star$ is the incident stellar flux and $\sigma$ is the Stefan-Boltzmann constant, varies from $500$ to $3000 \ \mathrm{K}$ in intervals of $500 \ \mathrm{K}$. Including the combined parameter space of equilibrium temperature and drag strength, this results in a grid of $36$ GCM simulations, with temperature and wind maps shown in Figure 5 of \cite{Komacek:2017}. 
\subsubsection{Tracer parameterization}
\label{sec:tracerparam}
In order to diagnose vertical tracer transport, we incorporate passive tracers into our GCM. We use a non-linear second-order flux limiter method to advect the passive tracers. This flux limiter method is used to improve the stability of the tracer scheme and the accuracy of the resulting tracer transport. We use a Shapiro filter to remove small scale grid noise without affecting large-scale physical structures. This filter is applied to the passive tracer field, potential temperature, and winds.  \\ 
\indent We utilize two separate types of tracer. The first of these tracers (the ``chemical relaxation tracer'') represents chemical species that react over a given reaction timescale $\tau_\mathrm{chem}$. This tracer setup can be considered to mimic the mixing of species that react due to chemical reactions over a given timescale $\tau_{\mathrm{chem}}$, and hence can be utilized to understand the impacts of vertical transport in hot Jupiter atmospheres on potential disequilibrium chemistry. This is implemented similarly to \cite{Cooper:2006}, who applied this scheme to mimic the specific case of CO/CH$_4$ interconversion in order to determine the effects of three-dimensional quenching on the relative abundances of CO and CH$_4$. This scheme is also similar to that used in the simulations of \cite{Zhang:2017} for tracers with uniform chemical timescale and chemical equilibrium mixing ratio. Recently, \cite{Tsai:2017} found that though predictions with such a chemical relaxation scheme are limited by the assumption of a single rate-limiting step, this simplified scheme is in agreement with more detailed chemical-kinetic schemes at the order-of-magnitude level.     \\
\indent In this scheme, the source/sink of tracer abundance is set to relax to a specified equilibrium source over a chemical timescale as
\begin{equation}
\label{eq:chemtrac}
\frac{dq}{dt} = -\frac{q - q_\mathrm{eq}}{\tau_\mathrm{chem}} \mathrm{.}
\end{equation}
In \Eq{eq:chemtrac}, $q$ is the tracer abundance and the material derivative is $d/dt = \partial/\partial t + {\bf v} \cdot \nabla + \omega \partial/\partial p$, where ${\bf v}$ is the horizontal velocity, $\nabla$ is the horizontal gradient on isobars, $\omega = dp/dt$ is the vertical velocity in pressure coordinates, and $p$ is pressure. The equilibrium tracer profile is taken to be a simple function in which $q_\mathrm{eq}$ decreases smoothly with decreasing pressure from a fixed abundance, $q_\mathrm{bot}$, at a reference pressure $p_\mathrm{bot}$:
\begin{equation}
q_\mathrm{eq} = \begin{cases} q_{\mathrm{bot}} \left(\frac{p}{p_{\mathrm{bot}}}\right)^{\zeta} \ p < p_\mathrm{bot}\mathrm{,} \\ q_\mathrm{bot} \hspace{1.37cm} p \ge p_\mathrm{bot} \mathrm{.} \end{cases}
\end{equation}
We take $q_{\mathrm{bot}} = 10^{-5}$ to be fixed, $p_\mathrm{bot}$ to be $400 \ \mathrm{mbar}$ (which as we will show lies near the bottom of the region with fast vertical velocities), and $\zeta = \mathrm{ln}(q_{\mathrm{top}}/q_{\mathrm{bot}})/\mathrm{ln}(p_{\mathrm{top}}/p_{\mathrm{bot}})$, where $q_{\mathrm{top}}$ is set to be small ($10^{-12}$) and $p_\mathrm{top} = 10^{-2} \ \mathrm{mbar}$. We expect that the values of $\Kzz$ should not depend sensitively on the exact values of $q_{\mathrm{bot}}$ and $q_{\mathrm{top}}$, as long as they are significantly different from each other; the main purpose is to produce a background vertical gradient that can be advected by the flow. \\
\indent In this work, we vary $\tau_{\mathrm{chem}}$ (which is taken to be a constant in space) from $10^4 - 10^6 \ \mathrm{s}$, evenly spaced in the base-10 logarithm (i.e. resulting in $\tau_\mathrm{chem}$ values of $10^4, 10^{4.4}, 10^{4.8}, 10^{5.2}, 10^{5.6}, 10^6 \ \mathrm{s}$). Although chemical timescales usually vary strongly with height, and likewise could vary strongly from dayside to nightside on a planet with a large day-night temperature difference, here for simplicity we assume that, in any given model, $\tau_\mathrm{chem}$ is constant both horizontally and vertically. However, in a real atmosphere the chemical timescale for CO/CH$_4$ interconversion would increase by orders of magnitude with decreasing pressure, from $\sim 10^5 \ \mathrm{s}$ at 10 bars to $\sim 10^{12} \ \mathrm{s}$ at 10 mbar for a planet with $T_\mathrm{eq} = 1500 \ \mathrm{K}$ \citep{Cooper:2006,Tsai:2017}. As a result, species are more likely to be in disequilibrium at lower pressures due to the increasing chemical timescale with decreasing pressure. \\
\indent In this work, we use a uniform chemical timescale that represents the deep timescale. We do so because the deep chemical timescale represents the chemical timescale at the quench level. The chemical timescale above the quench level does not greatly affect the tracer abundance, as the chemistry is quenched there. As we use a fixed chemical timescale with height, the mixing ratios of chemical species are closer to chemical equilibrium at lower pressures than they would be when including the effect of varying chemical timescale with pressure. This assumption of a uniform chemical timescale allows us to more readily characterize how the mixing properties scale with $\tau_\mathrm{chem}$. Additionally, this uniform chemical timescale promotes an easier comparison with analytic theory, which is more straightforward to develop for the case with $\tau_\mathrm{chem}$ taken to be a constant.    \\
\indent The second type of tracer we implement (the ``aerosol tracer'') represents particles that in the absence of dynamics settle at their terminal velocity through the atmosphere. This is exactly the same as the tracer scheme of \cite{parmentier_2013}, except here we take into account settling at all longitudes, while they only allowed settling on the nightside. Similarly to \cite{parmentier_2013}, we assume that settling happens only at pressures less than or equal to $1 \ \mathrm{bar}$, as we are interested in the transport of aerosols in observable regions of the atmosphere and to aid with numerical convergence of our simulations. We also use a Newtonian source/sink term with a relaxation timescale $\tau_\mathrm{relax} = 10^6~\mathrm{s}$ to keep the deep abundance at pressures $\> 1~\mathrm{bar}$ close to a prescribed value $q_\mathrm{eq,deep}$ which is equal to unity, as follows:
\begin{equation}
\frac{dq}{dt} = -\frac{q - q_\mathrm{eq,deep}}{\tau_\mathrm{relax}}  \hspace{0.5cm} (p > 1 \ \mathrm{bar}) \mathrm{.}
\end{equation}
This settling scheme mimics well the vertical transport of aerosols, though it does not take into account growth of particles. Instead, we use a fixed particle radius, and vary its value from $10^{-2} - 10 \mu \mathrm{m}$ (specifically, we use spherical particles of radius $0.01, 0.1, 0.5, 1, 2.5, 5, 10~\mu \mathrm{m}$, similar to \citealp{parmentier_2013}). All particles are assumed to have the same density of $4500 \ \mathrm{kg} \ \mathrm{m}^{-3}$, which lies between the densities of silicates and iron and titanium oxides. \\
\indent Formally, the source/sink of tracer due to settling in this model is
\begin{equation}
\label{eq:settletrac}
\frac{dq}{dt} = \frac{1}{\rho} \frac{\partial\left(\rho q V_{\mathrm{term}}\right)}{\partial z}  \hspace{0.5cm} (p \le 1 \ \mathrm{bar})\mathrm{,}
\end{equation}
where $\rho$ is the density of the surrounding air, $z$ is height, and $V_\mathrm{term}$ is the terminal velocity of the particles, given by
\begin{equation}
V_\mathrm{term} = \frac{2\beta r^2 g(\rho_{\mathrm{p}} - \rho)}{9\eta} \mathrm{,}
\end{equation}
where $r$ and $\rho_{\mathrm{p}}$ are the particle radius and density, respectively. $\beta$ is the Cunningham factor, which becomes important when the Knudsen number, $\mathrm{Kn} = \lambda/r$, where $\lambda$ is the mean free path in the atmosphere, becomes large. We use the same parameterization for the Cunningham factor as \cite{Spiegel:2009} and \cite{parmentier_2013}:
\begin{equation}
\beta = 1 + \mathrm{Kn}\left(1.256 + 0.4 e^{-1.1/\mathrm{Kn}}\right) \mathrm{.}
\end{equation}
Lastly, we assume the same parameterization of the viscosity of hydrogen from \cite{Rosner:2000} as \cite{Ackerman2001} and \cite{parmentier_2013}
\begin{equation}
\eta = \frac{5\sqrt{\pi m k_\mathrm{B}T}}{16 \pi d^2} \frac{\left(k_\mathrm{B}T/\epsilon\right)^{0.16}}{1.22} \mathrm{,}
\end{equation}
where $d = 2.827 \times 10^{-10} \ \mathrm{m}$ is the molecular diameter of $\mathrm{H}_2$, $m = 3.34 \times 10^{-27} \ \mathrm{kg}$ is the molecular mass of $\mathrm{H}_2$, and $\epsilon = 59.7k_\mathrm{B} \ \mathrm{K}$ is the depth of the Lennard-Jones potential well for $\mathrm{H}_2$. Note that the parameterization used for $\eta$ is valid at temperatures below $3000 \ \mathrm{K}$ and pressures below $100 \ \mathrm{bars}$.
However, at temperatures above $3000 \ \mathrm{K}$ hydrogen becomes partially ionized, making viscosity less dependent on temperature. Note we do extrapolate this formulation for our hottest runs, which have equilibrium temperatures of $2500$ and $3000 \ \mathrm{K}$ and hence can have local temperatures greater than $3000 \ \mathrm{K}$. 
\subsubsection{Numerical details}
We adopt planetary parameters relevant for HD 209458b for the vast majority of our simulations\footnote{Specifically, we take the specific heat $c_p = 1.3 \times 10^4 \mathrm{J} \mathrm{kg}^{-1} \mathrm{K}^{-1}$, specific gas constant $R = 3700 \ \mathrm{J} \mathrm{kg}^{-1} \mathrm{K}^{-1}$, planetary radius $a = 9.437 \times 10^7 \ \mathrm{m}$, gravity $g = 9.36 \ \mathrm{m}\mathrm{s}^{-2}$, and rotation rate $\Omega = 2.078 \times 10^{-5} \ \mathrm{s}^{-1}$.}, the same as in \cite{Komacek:2015} and \cite{Komacek:2017}. In \Sec{sec:theorycompkzz}, we will explore the effects of consistently varying rotation rate and equilibrium temperature on vertical mixing rates. We solve the equations of motion on the full sphere with a cubed sphere grid. As in \cite{Komacek:2015} and \cite{Komacek:2017}, we use a horizontal resolution of C32 (which roughly corresponds to a global resolution of $128 \times 64$ in longitude and latitude, respectively) and 40 vertical pressure levels, with the bottom 39 spaced evenly in log-pressure between $200 \ \mathrm{bars}$ (the bottom of the domain) and $0.2 \ \mathrm{mbars}$, with a top layer that extends to zero pressure. We integrate our models to $2,000 \ \mathrm{Earth} \ \mathrm{days}$. All simulations except the weakly forced simulations with $T_\mathrm{eq} = 500 \ \mathrm{K}$ and weakly damped simulations with $\tau_\mathrm{drag} = \infty$ reach a steady-state in domain-integrated kinetic energy by 2,000 days of model time. All results that are shown below are time-averaged over the last $100 \ \mathrm{days}$ of model time. 
\subsection{Vertical velocities}
\label{sec:vertvel}
\begin{figure*}
	\centering
	\includegraphics[width=1\textwidth]{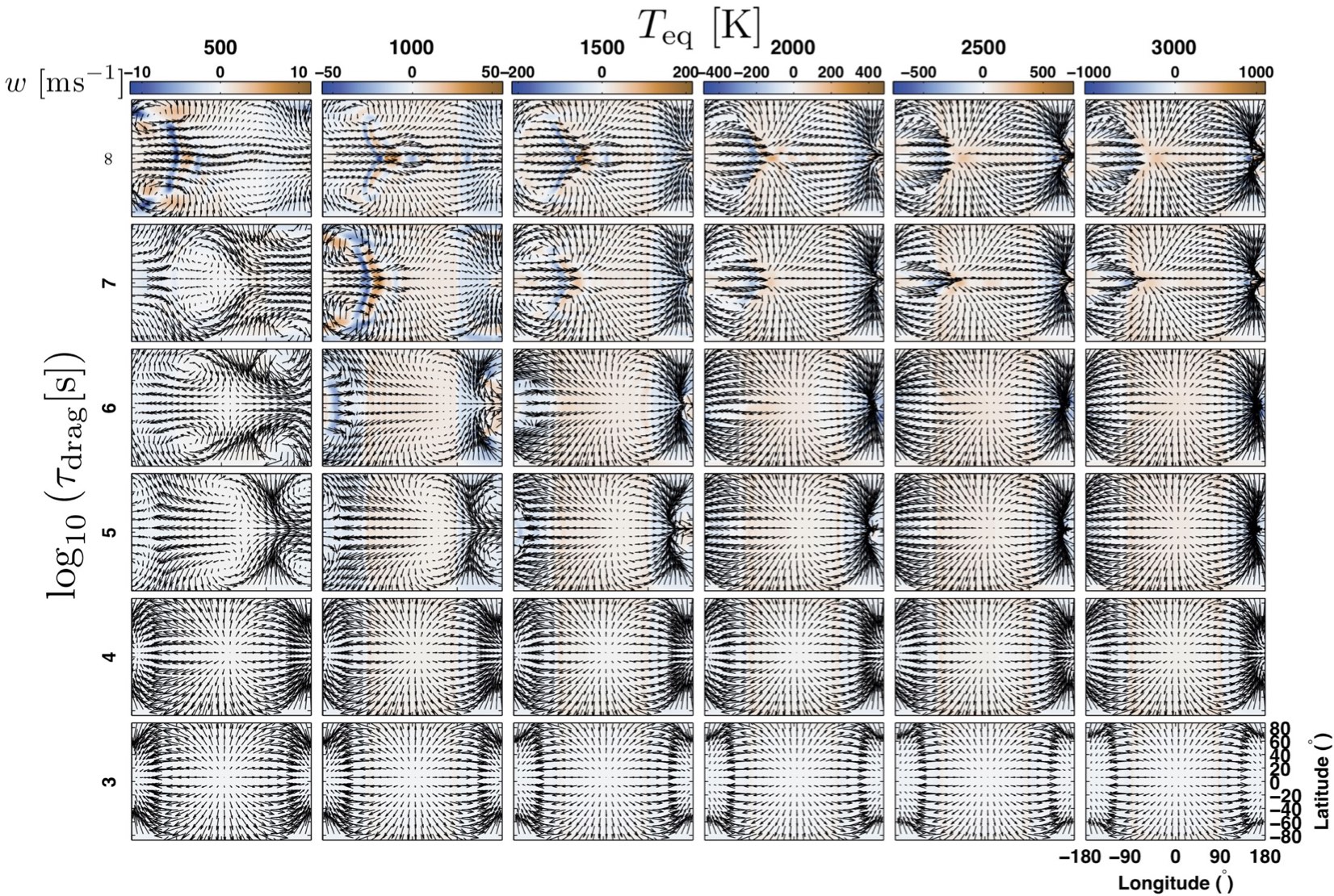}
	\caption{Maps of vertical velocity (colors, in $\mathrm{m}\mathrm{s}^{-1}$) and winds (arrows, lengths represent wind speed) at $1 \ \mathrm{mbar}$ pressure. Results are from 36 separate GCM simulations varying the incident stellar flux (corresponding to global-mean equilibrium temperatures of $500-3000 \ \mathrm{K}$) and drag timescale (from $10^3 \ \mathrm{s} - \infty$). Each column shares a color scale, with redder (bluer) colors corresponding to faster upward (downard) vertical velocities. The longitude and latitude scales are identical on each of the 36 panels, but for clarity are only displayed on the lower-right panel. The substellar point lies in the middle of each panel, i.e., at a longitude and latitude of zero. Note the transition in the flow between the chevron-like patterns associated with a superrotating jet at weak drag ($\tau_{\mathrm{drag}} \ge 10^7 \ \mathrm{s}$) to day-to-night flow at $\tau_\mathrm{drag} \le 10^6 \ \mathrm{s}$.}
	\label{fig:wmap}
\end{figure*}
\begin{figure*}
	\centering
	\includegraphics[width=1\textwidth]{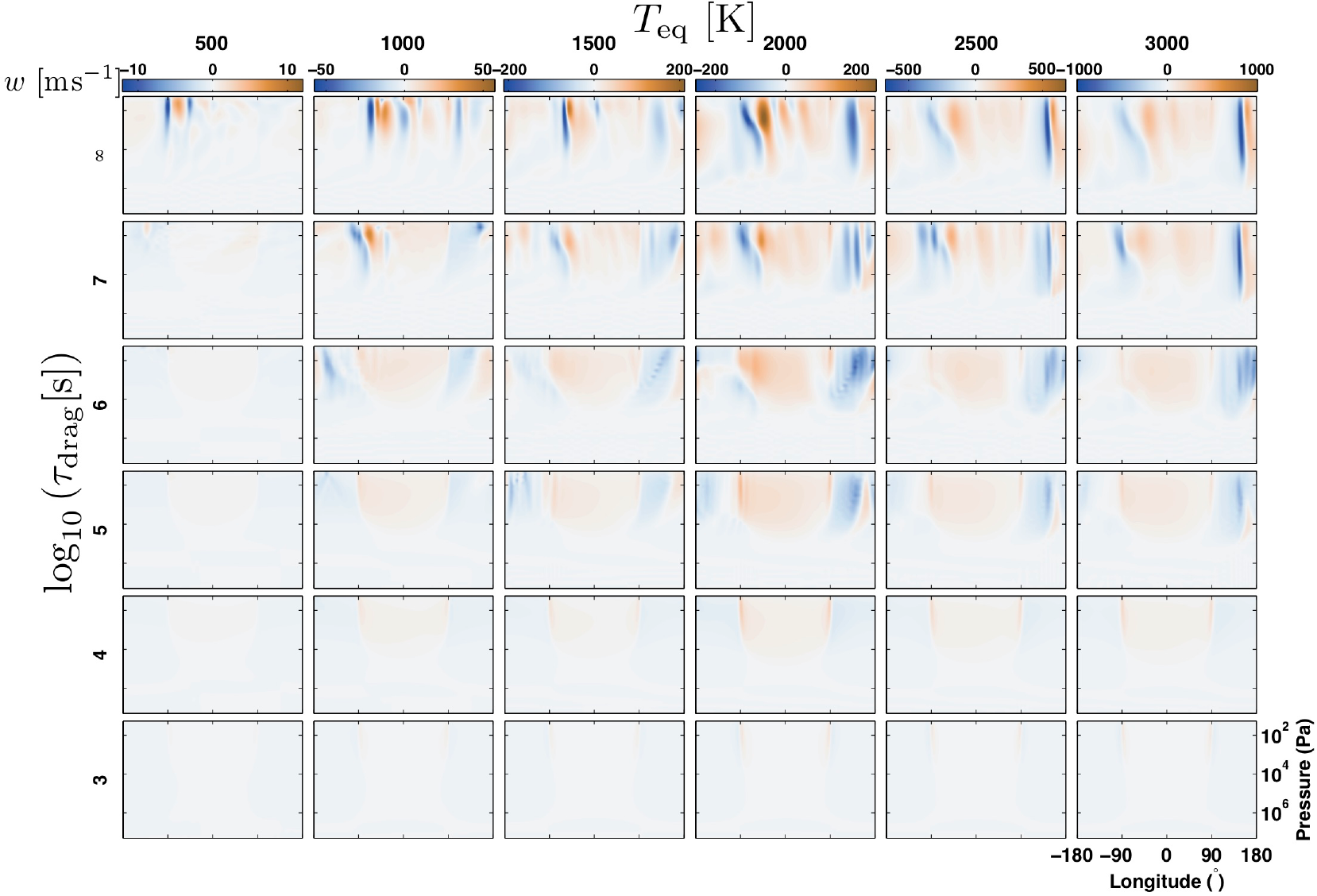}
	\caption{Equatorial slices of vertical velocity (colors, in $\mathrm{m}\mathrm{s}^{-1}$) from 36 separate GCM simulations varying the incident stellar flux (parameterized by equilibrium temperature) and frictional drag timescale. Each column shares a color scale, with redder (bluer) colors corresponding to faster upward (downard) vertical velocities. The longitude and pressure scales are identical on each of the 36 panels, but for clarity are only displayed on the lower-right panel. When drag is weak, with characteristic $\tau_{\mathrm{drag}} \ge 10^7 \ \mathrm{s}$, the vertical velocities exhibit a columnar structure, with the fastest vertical velocities at places where the superrotating jet increases or decreases in speed. There is a transition with stronger drag ($\tau_{\mathrm{drag}} \le 10^6 \ \mathrm{s}$) to a flow that is instead characterized by upwelling everywhere on the dayside and downwelling throughout the nightside.} 
	\label{fig:weq}
\end{figure*}
Before analyzing the mixing of passive tracers, we first examine the nature of the vertical flow in our simulated hot Jupiter atmospheres. \Fig{fig:wmap} shows maps of vertical velocity at $1 \ \mathrm{mbar}$ pressure from our grid of 36 GCM experiments, while \Fig{fig:weq} shows equatorial slices of the vertical velocity for each of these models. Note that here we plot the vertical velocity in log-pressure coordinates, $w = -Hd\mathrm{ln}p/dt$, where $H(p) = RT(p)/g$ is the atmospheric scale height that varies with pressure $p$. \\
\indent Examining \Fig{fig:wmap}, one can see that our simulations with weak drag ($\tau_{\mathrm{drag}} \ge 10^7 \ \mathrm{s}$) have a superrotating jet at the equator and resulting localized regions with strongly positive and strongly negative vertical velocity at locations where this superrotating jet abruptly increases or decreases in speed. These features are most prominent near the western limb of the planet, and manifest as a chevron-like pattern in the vertical velocity. Meanwhile, if drag is relatively strong ($\tau_{\mathrm{drag}} \le 10^6 \ \mathrm{s}$), there is no superrotating jet and the vertical velocity is positive everywhere on the dayside and negative throughout much of the nightside. \\
\indent The effects of the characteristic transition in the flow from superrotating to day-night flow (as found previously in \citealp{showman_2013_doppler}) on the vertical velocity is clearer when examining \Fig{fig:weq}. In the superrotating regime, the features where there is convergence/divergence driven by the equatorial jet changing in speed manifest as columns of vertical velocity which are coherent between pressures of $\sim 0.1-10 \ \mathrm{bar}$. This general character of the flow when drag is weak agrees well with that studied in \cite{parmentier_2013}, who examined vertical transport for the specific case of HD 209458b (see their Figure 3). However, we have found that when including the effects of drag, the lack of an eastward equatorial jet leads to a drastically different character of the vertical flow. Instead of being columnar, the flow is from day-to-night, with corresponding upwelling everywhere on the dayside and downwelling over much of the nightside. As we will see in \Sec{sec:tracer}, this change in the character of the flow from superrotating to day-night greatly affects vertical transport. 
\subsection{Tracer mixing ratios}
\label{sec:tracer}
\subsubsection{Chemical relaxation tracers}
\label{sec:tauchemtracer}
\begin{figure*}
	\centering
	\includegraphics[width=1\textwidth]{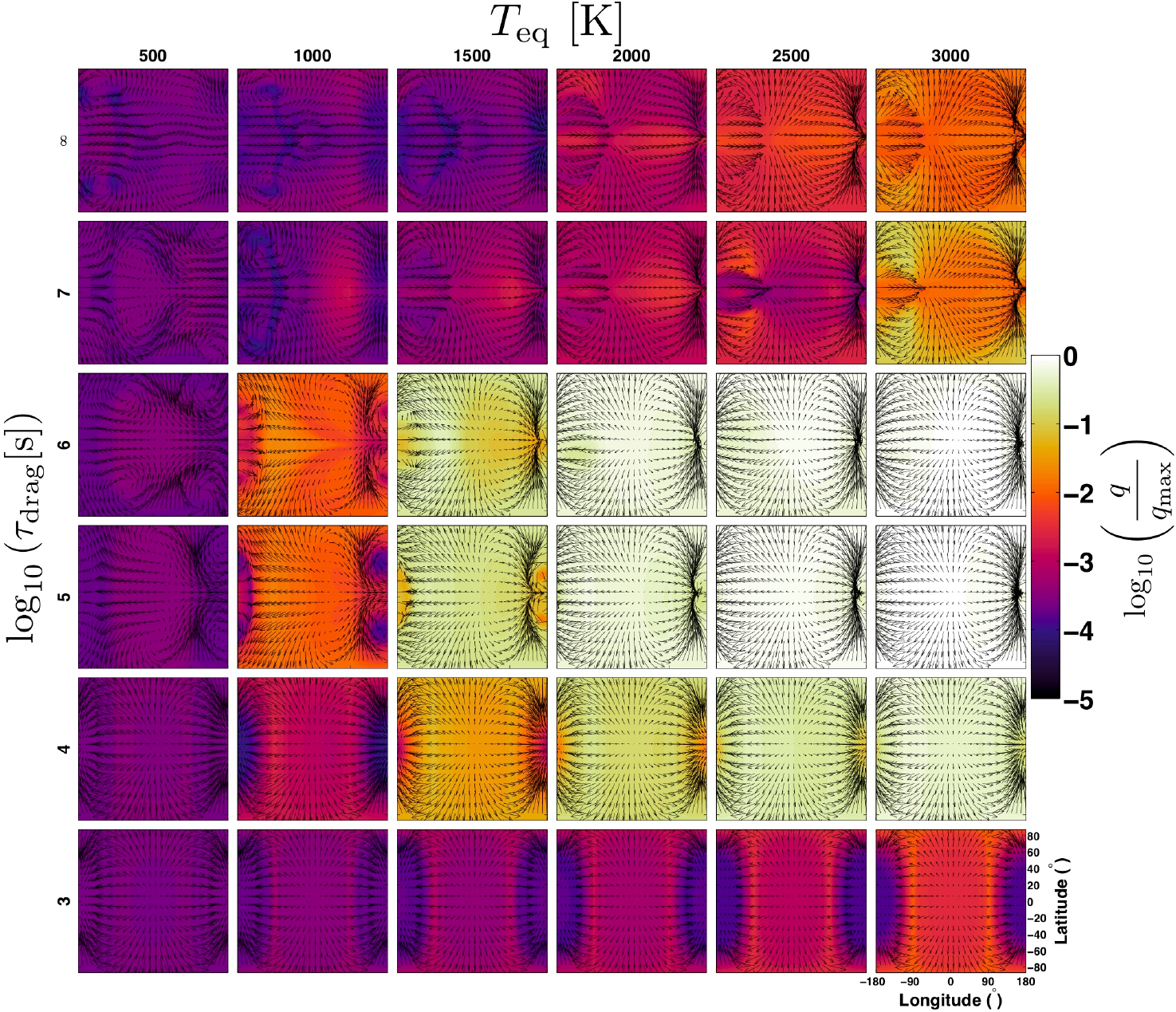}
	\caption{Maps of tracer abundance (colors, normalized to the maximum abundance in the entire set of models) from 36 separate GCM simulations including a passive chemical tracer with $\tau_\mathrm{chem} = 1.6 \times 10^5 \ \mathrm{s}$. The results are shown at $1 \ \mathrm{mbar}$ pressure as a function of incident stellar flux (parameterized by equilibrium temperature) and drag timescale. The longitude and latitude scales are identical on each of the 36 panels, but for clarity are only displayed on the lower-right panel. In the absence of drag, tracer abundance increases monotonically with increasing incident stellar flux. However, there is a local maximum in tracer abundance as a function of drag timescale for models with significant incident stellar flux, with orders-of-magnitude larger tracer abundance at intermediate drag strength ($\tau_\mathrm{drag} \sim 10^5-10^6$ s) than models with either weaker or stronger drag. This is due to the transition of the flow from dominated by a superrotating jet (which is relatively poor at vertical transport) with long $\tau_\mathrm{drag} \ge 10^7 \ \mathrm{s}$ to day-to-night flow with shorter $\tau_\mathrm{drag} \le 10^6 \ \mathrm{s}$.}
	\label{fig:tauchem}
\end{figure*}
To show how the transition in the character of the flow from superrotating to day-to-night discussed in \Sec{sec:vertvel} affects tracer mixing, we examine chemical tracers (with a source/sink prescribed by Equation \ref{eq:chemtrac}). We focus here on describing the results from one intermediate chemical timescale ($1.6 \times 10^5 \ \mathrm{s}$). \Fig{fig:tauchem} shows the resulting tracer maps at a pressure of $1 \ \mathrm{mbar}$ with varying equilibrium temperature and drag strength. The transition from superrotation to day-to-night flow as the drag timescale decreases from $10^7~\mathrm{s}$ to $10^6~\mathrm{s}$ causes a large increase in the tracer mixing ratio, by two to three orders of magnitude. Then, when drag becomes even stronger in the day-to-night flow regime ($\tau_\mathrm{drag} \lesssim 10^4~\mathrm{s}$), the flow is damped enough such that the tracer mixing ratios again become small. Note that we also find this maximum in tracer abundance at intermediate $\tau_\mathrm{drag}$ for all the values of $\tau_\mathrm{chem}$ that we explored, and for the mixing of aerosol tracers. The behavior of decreasing tracer mixing ratio with increasing drag strength seen in the bottom half of \Fig{fig:tauchem} is expected, because increasing the drag strength damps the circulation and causes the wind speeds to decrease, particularly when $\tau_\mathrm{drag} \lesssim 10^4~\mathrm{s}$. The circulation therefore advects the tracer more slowly, and thus -- for a given $\tau_\mathrm{chem}$ -- it is easier for the chemistry to relax the tracer abundance to a value closer to equilibrium when the drag timescale is shorter. In contrast, the local maximum of large tracer abundance at intermediate drag strength and hot equilibrium temperatures is puzzling and requires deeper investigation. \\
\indent To understand the peak in tracer mixing ratio at intermediate $\tau_\mathrm{drag}$, we turn back to the pattern of vertical velocity shown in \Fig{fig:weq}. As discussed earlier, in the superrotating regime there are neighboring columns of upward/downward vertical velocity, while in the day-to-night flow regime the entire dayside is upwelling and most of the nightside is downwelling. If one envisions a particle released on the dayside in the middle of the superrotating jet, it will on short timescales be advected eastward through the ``chimneys'' of upward and downward velocities shown in Figures 1-2 (which are quasi-steady in their spatial positions). Due to the quickly varying vertical velocity from positive to negative as the particle travels eastward it would only have small excursions from its initial pressure. Moreover, as the air parcel travels eastward across any given ``chimney'' of high vertical velocity, the vertical {\it displacement} from some reference pressure will essentially be an integral of the vertical velocity following the air parcel. Because it takes time for the parcel to move vertically by finite displacements, the minima/maxima of the air parcel's vertical displacement will be phase shifted to the east relative to the minima/maxima of the velocity field itself. Given a background vertical gradient in tracer abundance, the tracer anomalies on an isobar should correlate spatially with the displacement field (as long as $\tau_\mathrm{chem}$ is not so long as to allow the mixing to destroy the background vertical tracer gradient). Therefore, the minima/maxima of tracer abundance should be shifted eastward of the vertical velocity field in the situation with a strong superrotating jet.  This leads to a correlation between the tracer field and the vertical velocity field that is substantially less than 1\footnote{One could formally define such a correlation coefficient, for example, as $C=\langle q w\rangle/(q_{rms} w_{rms})$, where $q_{rms}$ and $w_\mathrm{rms}$ are the root-mean-square values of tracer anomaly (that is, the deviation of the tracer from its horizontal average) and of the vertical velocity on a given isobar, calculated over the globe. This correlation coefficient would be close to 1 when the spatial variations of $q$ and $w$ on an isobar are perfectly correlated and 0 when they are uncorrelated.}. It is this relatively weak correlation coefficient between the horizontal pattern of vertical velocity and tracer abundance in the superrotating regime that leads to relatively weak vertical mixing rates and therefore a smaller tracer abundance at the longest drag timescales in Figure 3 (relative to $\tau_\mathrm{drag} \sim 10^5-10^6$ s).  \\
\indent In contrast, if a particle is released on the dayside in the day-to-night flow regime, it would continually move upward as it is travels horizontally, as it does not encounter downwelling regions until it reaches the nightside. More importantly, because the zonal-mean zonal wind is weak, there exists a much stronger correlation between the (horizontal variations of the) tracer field and the vertical velocity field for all values of $\tau_\mathrm{drag} \lesssim 10^6$ s. In general, for given approximate amplitudes of vertical velocity and tracer anomalies on isobars, it is much more efficient to mix species when the vertical velocity is spatially well correlated with the the deviation of tracer abundance on isobars. For example, mixing is more efficient if motions are upward where there is a relatively high tracer abundance and downward where there is a relatively low tracer abundance \citep{Zhang:2017}. \\
\indent This coherence between vertical velocity and tracer abundance is best in our simulations of hot Jupiters with strong $\tau_\mathrm{drag} \le 10^6 \ \mathrm{s}$, leading to large effective vertical mixing rates. This explains why the tracer abundances increase with increasing drag strength between $\tau_\mathrm{drag} = \infty$ (no drag) and $\tau_\mathrm{drag} \sim 10^5$ s for $T_\mathrm{eq} \gtrsim 1000$ K.  For further increases in drag strength (moving downard on \Fig{fig:tauchem}), the coherence between tracer abundance and vertical velocity pattern remains high, but the drag becomes so strong that it starts to significantly damp the wind speeds, which lessens the mixing rate and therefore lessens the tracer abundance at high altitude.  Together, these explain the maximum in tracer abundance as a function of $\tau_\mathrm{drag}$ in \Fig{fig:tauchem}. Essentially, the local maximum in tracer abundance occurs because of a happy medium where the amplitudes of vertical velocities and tracer anomalies are high and also are well correlated; at higher or lower $\tau_\mathrm{drag}$, at least one of the conditions for this happy medium are not satisfied.
\begin{figure*}
	\centering
	\includegraphics[width=1\textwidth]{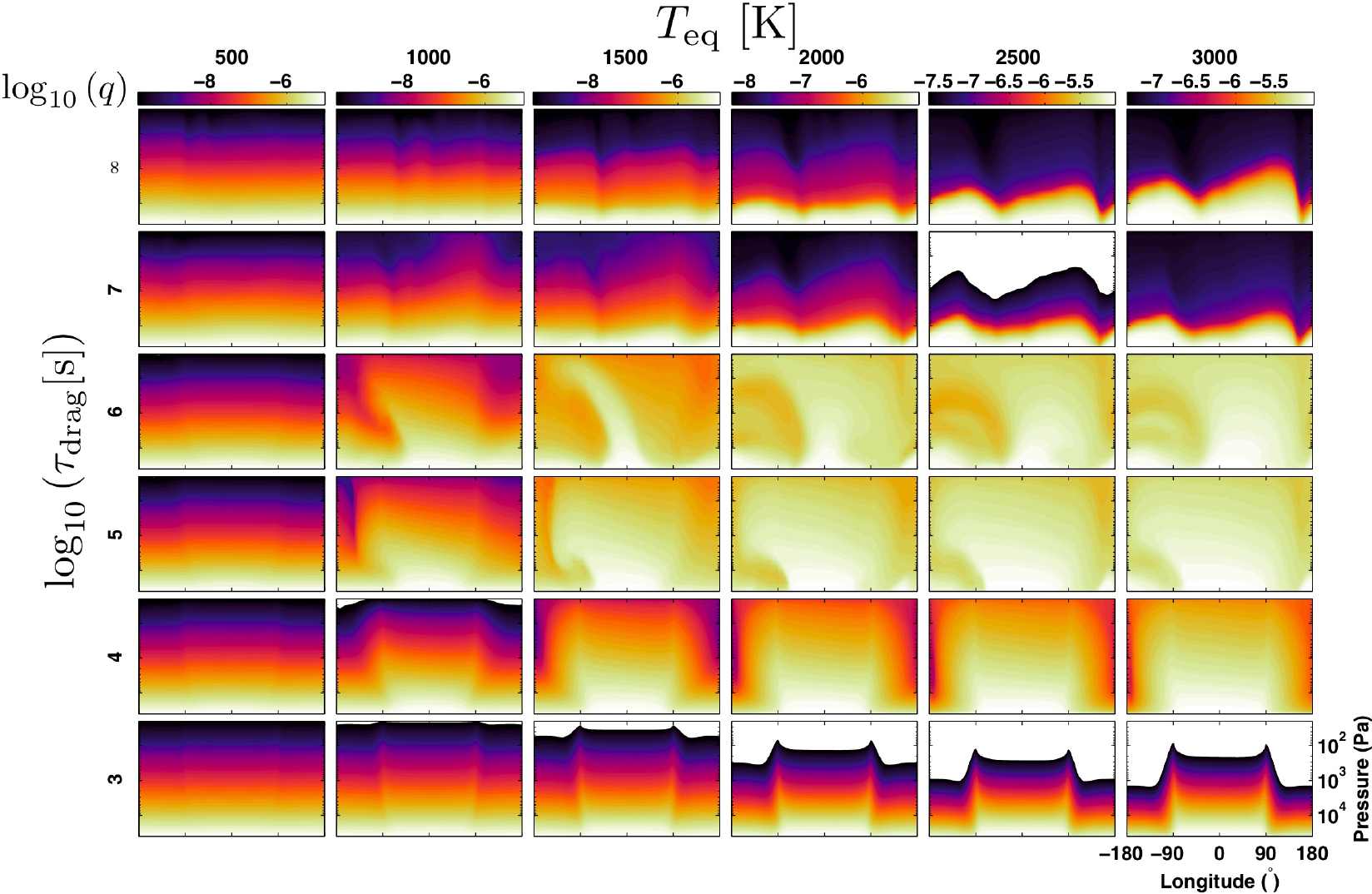}
	\caption{Contours of tracer abundance at the equator (colors, showing the $\mathrm{log}_{10}$ of absolute tracer abundance) as a function of pressure (ordinate) and longitude (abcissa). The tracer chemical relaxation timescale is $\tau_\mathrm{chem} = 1.6 \times 10^5 \ \mathrm{s}$, and results are shown for simulations varying the incident stellar flux (parameterized by equilibrium temperature) and drag timescale. The longitude and pressure scales are identical on each of the 36 panels, but for clarity are only displayed on the lower-right panel. Lighter colors correspond to larger tracer abundances, and white regions above black regions correspond to levels where $q < 10^{-10}$. We only show these maps at pressures less than $4 \times 10^4 \ \mathrm{Pa}$, as the equilibrium tracer abundance is set to a large fixed value at greater pressures. As shown in \Fig{fig:tauchem}, the tracer abundance at low pressures is largest for simulations with intermediate $10^4 \ \mathrm{s} \lesssim \tau_\mathrm{drag} \lesssim 10^6 \ \mathrm{s}$. Additionally, the tracer abundance is largest on the dayside, with advection onto the nightside occurring at low pressures where horizontal winds are strongest.}
	\label{fig:tauchem_pressure}
\end{figure*}
\\ \indent Our result from \Fig{fig:tauchem} that chemical tracer abundance reaches a local maximum at intermediate $\tau_\mathrm{drag}$ holds well over a broad range of pressures. \Fig{fig:tauchem_pressure} shows the equatorial tracer abundance as a function of pressure and longitude from the same set of simulations as shown in \Fig{fig:tauchem}. Comparing the tracer abundances from \Fig{fig:tauchem_pressure} with the vertical velocities from \Fig{fig:weq}, one can see that the peak in tracer abundance at intermediate $10^4 \ \mathrm{s} \lesssim \tau_\mathrm{drag} \lesssim 10^6 \ \mathrm{s}$ is associated with a horizontal coherence between upwelling motions and tracer abundance. We find that the increase in tracer abundance with intermediate $\tau_\mathrm{drag}$ continues up to very low pressures, $\sim 0.1 \ \mathrm{mbar}$. Meanwhile, at both larger and smaller $\tau_\mathrm{drag}$ there is a slight increase in the tracer abundance at lower pressures, but the vertical transport is not strong enough to mix species to low pressures.  \\
\indent In our simulations with intermediate $\tau_\mathrm{drag}$, the tracer abundance is largest on the dayside of the planet, as vertical velocities are upward throughout the dayside. When drag is strong, the maximum in tracer abundance in longitude occurs at the substellar longitude of zero and the overall pattern is nearly symmetric in longitude about this substellar longitude (\Fig{fig:tauchem_pressure}, bottom two rows).  On the other hand, when drag is weak and a superrotating jet exists, the maxima in tracer are shifted away from the substellar longitude (\Fig{fig:tauchem_pressure}, top rows). 
\begin{figure*}
	\centering
	\includegraphics[width=1\textwidth]{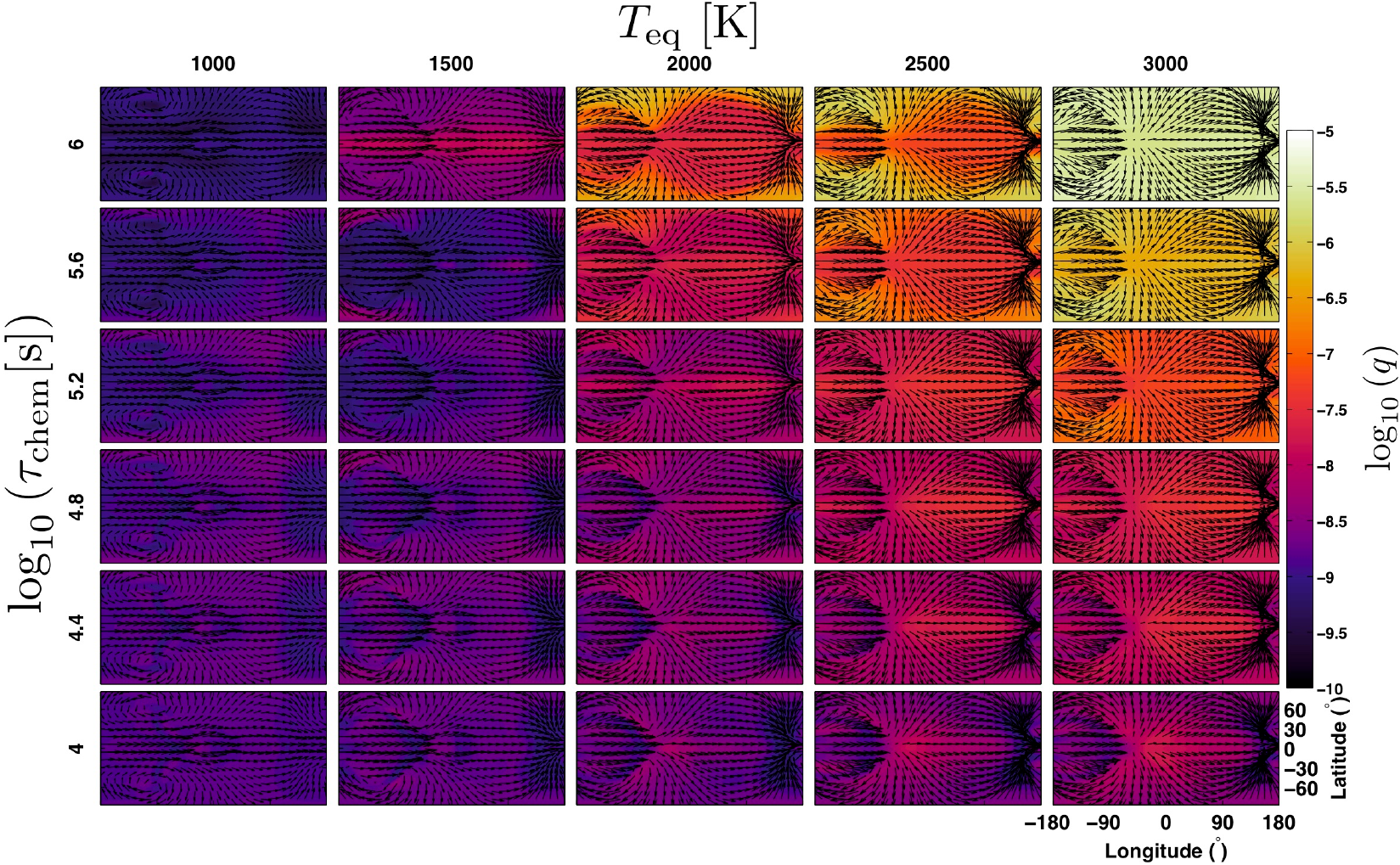}
	\caption{Maps of tracer abundance (colors, here showing the $\mathrm{log}_{10}$ of absolute tracer abundance) using a passive chemical relaxation tracer with varying $\tau_\mathrm{chem}$ and incident stellar flux (parameterized by equilibrium temperature). Results are shown at $1~\mathrm{mbar}$ pressure from GCM simulations with no applied drag at pressures less than $10~\mathrm{bar}$. The longitude and latitude scales are identical on each of the 36 panels, but for clarity are only displayed on the lower-right panel. In general, the tracer abundances at $1~\mathrm{mbar}$ increase with increasing incident stellar flux and increasing chemical relaxation timescale. This increase in tracer abundance can be as large as a factor of $\sim 10^3$  when varying the chemical relaxation timescale from $10^4~\mathrm{s}$ to $10^6~\mathrm{s}$. Models that receive more incident stellar flux show a larger change in the tracer abundance with varying chemical timescale, as hotter planets have more vigorous atmospheric vertical transport.}
	\label{fig:tauchemteq}
\end{figure*}
\begin{figure*}
	\centering
	\includegraphics[width=1\textwidth]{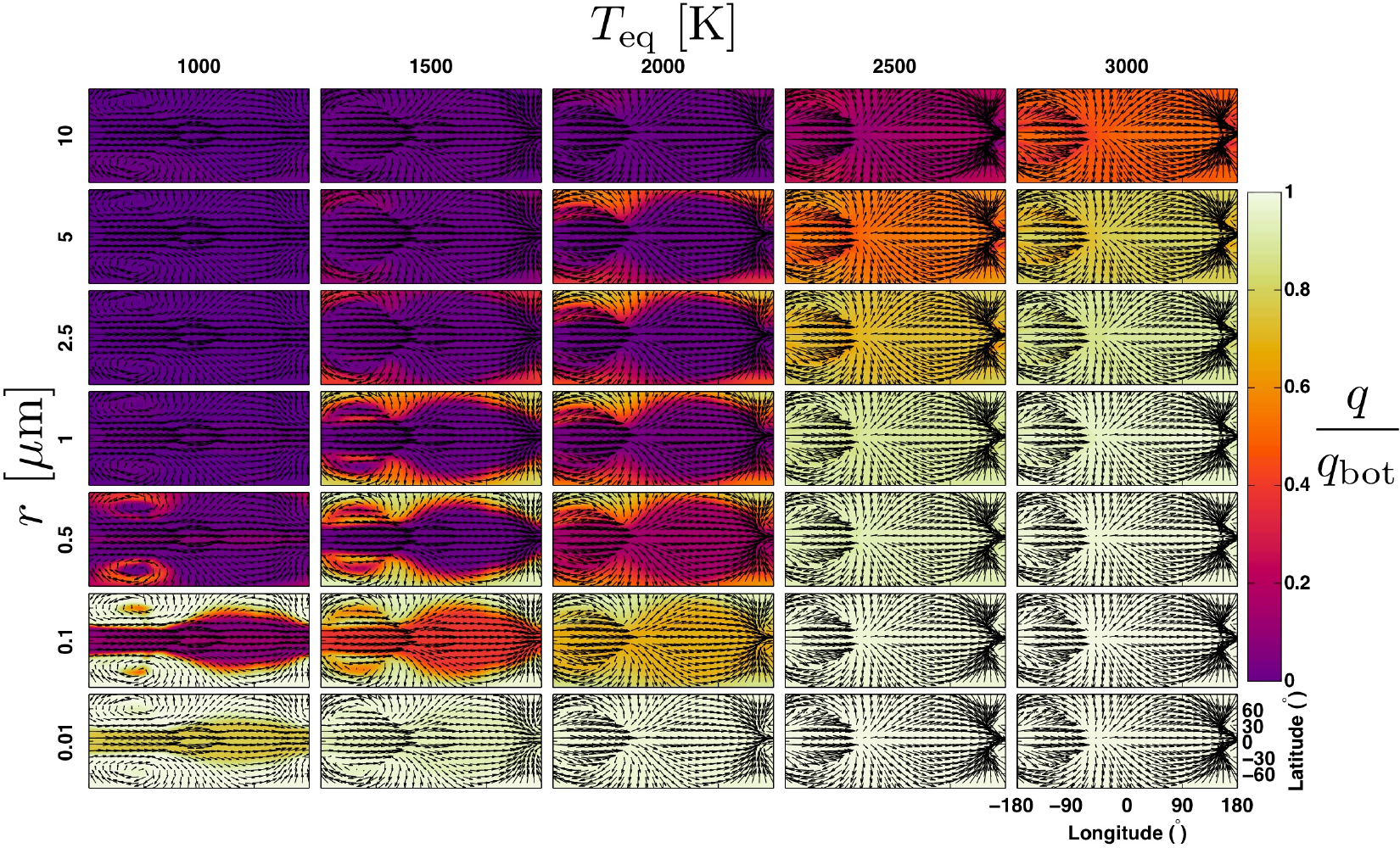}
	\caption{Maps of tracer abundance (colors, normalized to the deep abundance) from GCM simulations that adopt a passive aerosol tracer with varying particle radius from $0.01 - 10~\mu\mathrm{m}$. The tracer abundance is shown at a pressure of $1~\mathrm{mbar}$ from simulations varying particle size and varying incident stellar flux (parameterized by equilibrium temperature), with $\tau_\mathrm{drag} = \infty$. Generally, simulations with a larger incident stellar flux are more efficient at lofting passive aerosol tracers vertically. Additionally, tracers with smaller particle sizes are more efficiently lofted. As seen in \cite{parmentier_2013} and \cite{Lines:2018}, aerosol tracers are most depleted in the equatorial regions, with the strongest equatorial depletion occurring for cooler planets ($T_\mathrm{eq} \le 2000$ K) and particle sizes of $0.1-5$ microns, depending on the exact effective temperature.}
	\label{fig:settling_1mbar}
\end{figure*}
\indent Next we turn to the dependence of tracer mixing ratio on the tracer relaxation timescale, $\tau_\mathrm{chem}$. \Fig{fig:tauchemteq} shows how the tracer abundance varies with $T_\mathrm{eq}$ and chemical relaxation timescale for simulations with no drag in the free atmosphere\footnote{Note that all simulations contain a frictional drag scheme near the base of the model at pressures greater than 10 bars.}.  Essentially, \Fig{fig:tauchemteq} shows how the tracer abundance changes for the 3D dynamical flows depicted in the top row of Figure 3, but when different values of $\tau_\mathrm{chem}$ are used. As shown in \Fig{fig:tauchem}, the tracer abundance increases with increasing equilibrium temperature. Additionally, we show in \Fig{fig:tauchemteq} that the tracer abundance increases with increasing chemical timescale. That is, tracers with longer relaxation timescales take longer to return to their chemical equilibrium abundance, and as a result the circulation has more time over which it can act to mix the species vertically. Note that near the top of the model the difference between the abundance of two tracers, one with a long chemical timescale and and one with a short chemical timescale, increases with increasing equilibrium temperature. Aloft, the tracer abundance is only similar to the deep tracer abundance in chemical equilibrium ($q = 10^{-5}$) at the hottest equilibrium temperatures ($T_\mathrm{eq} = 3000~\mathrm{K}$) and longest chemical timescales ($\tau_\mathrm{chem} = 10^6~\mathrm{s}$) considered in our model grid. This is because simulations with larger incident stellar flux have faster vertical winds and faster vertical mixing rates. For all other simulations, there is significant (up to 5 orders of magnitude) depletion in the tracer abundance due to chemical relaxation.
\subsubsection{Aerosol tracers}
\begin{figure*}
	\centering
	\includegraphics[width=1\textwidth]{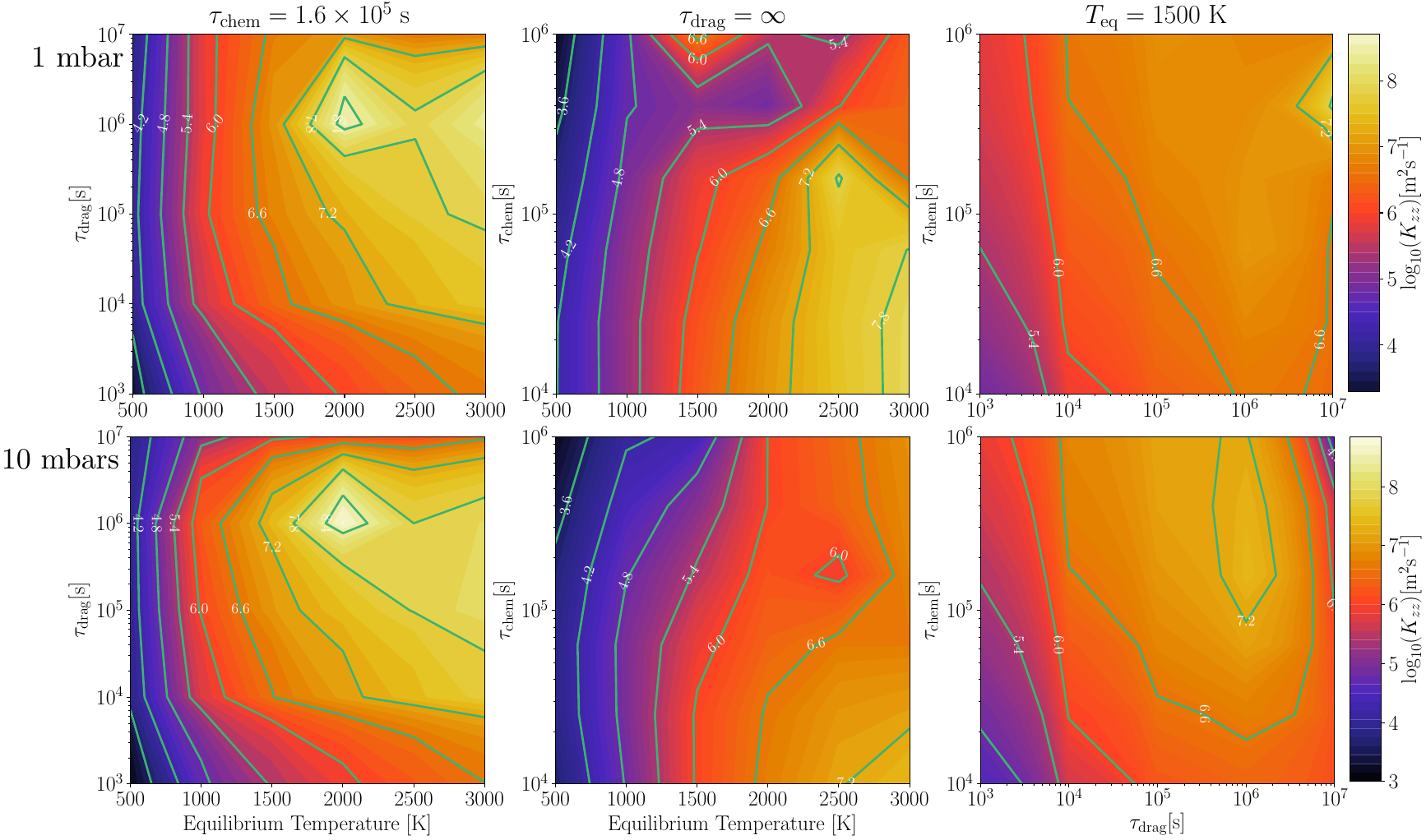}
	\caption{Vertical diffusivity $\Kzz$ at $1$ and $10 \ \mathrm{mbar}$ pressure calculated from our suite of GCM experiments using a chemical relaxation tracer. The figure shows three cuts through a 3D parameter space of $\Kzz$ as a function of $T_\mathrm{eq}$, $\tau_\mathrm{drag}$, and $\tau_\mathrm{chem}$. Respectively, the left, middle, and right plots show $\Kzz$ as a function of drag timescale and equilibrium temperature (with fixed $\tau_\mathrm{chem} = 1.6 \times 10^5~\mathrm{s}$), chemical timescale and equilibrium temperature (with fixed $\tau_{\mathrm{drag}} = \infty$), and chemical timescale and drag timescale (with fixed $T_\mathrm{eq} = 1500 \ \mathrm{K}$). The top row of plots shows $\Kzz$ at a pressure of $1 \ \mathrm{mbar}$, and the bottom row of plots shows $\Kzz$ at $10 \ \mathrm{mbars}$. Note that the left and middle plots can be viewed as $\Kzz$ versions of Figures \ref{fig:tauchem} and \ref{fig:tauchemteq}, respectively. Brighter colors correspond to larger $\Kzz$, and contours show $\mathrm{log}_{10}\Kzz$ in $\mathrm{m}^2\mathrm{s}^{-1}$. As expected, $\Kzz$ generally increases with increasing equilibrium temperature and increasing drag timescale. However, in agreement with our findings for the tracer mixing ratio itself, we find that $\Kzz$ does not show purely monotonic behavior with varying drag timescale. Additionally, there is non-monotonic behavior when varying the chemical loss timescale; note the local minimum in the middle figure when $\tau_\mathrm{chem} \sim 5 \times 10^5 \ \mathrm{s}$ and the local maximum in the right-hand figure at a similar value of $\tau_\mathrm{chem}$ and with $\tau_{\mathrm{drag}} = 10^7 \ \mathrm{s}$.}
	\label{fig:kzz_tauchem_drag_teq}
\end{figure*}
\indent Now we analyze the structure of passive aerosol tracers, relevant for understanding cloud and haze distributions in hot Jupiter atmospheres. Previously, \cite{parmentier_2013} used the same tracer parameterization as we applied in this work, except that they only allowed settling to occur on the nightside, whereas here we allow settling to occur at all longitudes. \citeauthor{parmentier_2013} found that tracers with smaller particle sizes were more efficiently mixed. However, because they only studied the vertical transport in the atmosphere of one specific hot Jupiter (HD 209458b), they did not analyze how the mixing of aerosol tracers depends on the incident stellar flux. Note that we find a similar local maximum in tracer abundance at intermediate $\tau_\mathrm{drag} \sim 10^6 \ \mathrm{s}$ in our simulations with an aerosol tracer and particle sizes $\ge 1 \ \mu\mathrm{m}$ as in our simulations with a chemical relaxation tracer. Presumably, the same transition from superrotation at longer $\tau_\mathrm{drag}$ to day-to-night flow at shorter $\tau_\mathrm{drag}$ that causes a local maximum in tracer mixing ratio for our chemical relaxation tracers also leads to this local maximum with varying $\tau_\mathrm{drag}$ for our aerosol tracers. We will return to the local maximum in vertical transport with varying drag timescale in \Sec{sec:kzz}. In this section, we focus on the dependence of tracer mixing ratio on particle size and incident stellar flux. \\
\indent \Fig{fig:settling_1mbar} shows how the aerosol tracer abundance depends on the particle size and incident stellar flux. 
We show results at a pressure of $1~\mathrm{mbar}$ for simulations with no applied drag (above the basal drag layer) and varying $T_\mathrm{eq} = 1000 - 3000~\mathrm{K}$ and particle size $r = 0.01 - 10~\mu\mathrm{m}$. We do not show simulations with $T_\mathrm{eq} = 500~\mathrm{K}$ because they only loft extremely small particle sizes ($\lesssim 0.1~\mu\mathrm{m}$) to low pressures. This is similar to Figure 6 of \cite{parmentier_2013}, but here we analyze how the mixing depends on both particle size and incident stellar flux, rather than simply particle size. \\
\indent We find that the tracer mixing ratio strongly increases with increasing incident stellar flux. This is because at larger values of incident stellar flux, the wind speeds are correspondingly stronger (see Figure 2). The tracer mixing ratio also increases with decreasing particle size, as smaller particles have a lower terminal velocity and hence are more easily lofted upward by the circulation. \\
\indent We find a depletion of tracer at the equatorial regions, most evident in our simulations with $T_\mathrm{eq} \le 2000~\mathrm{K}$ and particle sizes $\ge 0.1~\mu\mathrm{m}$. This equatorial depletion was also seen in the simulations of \cite{parmentier_2013}.  As in \cite{parmentier_2013}, we find that this equatorial depletion is not as strong for small particle sizes ($\le 0.1~\mu\mathrm{m}$). Moreover, this equatorial depletion does not occur at T=1000 K if the particle size exceeds $\sim 1 \ \mu\mathrm{m}$, and we additionally find that this equatorial depletion does not occur for simulations with hot $T_\mathrm{eq} \ge 2500~\mathrm{K}$. Additionally, the equatorial depletion was found by \cite{Lines:2018} using active cloud particles. 
Note that we also find a noticeable equatorial depletion in our runs with a chemical relaxation tracer that have long chemical timescales (see \Fig{fig:tauchemteq}), so the equatorial depletion is not an artifact of our aerosol tracer scheme. 
\subsection{Vertical mixing rates}
\label{sec:kzz}
\subsubsection{Calculation of $\Kzz$}
Now we examine how the vertical mixing rates (here parameterized as an effective diffusivity $\Kzz$) vary with the tracer sources/sinks, equilibrium temperature, and drag strength. This calculation does not depend on the method used for the tracers -- whether it be relaxation of the tracer toward a prescribed chemical equilibrium, or settling as would occur with falling particles -- we can use the same method to estimate $\Kzz$ from the simulations. We calculate $\Kzz$ from our GCM experiments as in \cite{parmentier_2013},
\begin{equation}
\label{eq:kzzcalc}
\Kzz = -\frac{\langle \rho q w\rangle}{\langle \rho \frac{\partial q}{\partial z} \rangle} \mathrm{,}
\end{equation}
where brackets represent an average on isobars across the entire planet. \Eq{eq:kzzcalc} can be derived from the flux-gradient relationship \citep{Plumb:1987aa}, and is generally applicable to situations with modest horizontal tracer anomalies relative to the vertical variations in tracer, such that there exists a well-defined vertical gradient of the background tracer field.  Note that \Eq{eq:kzzcalc} is applicable regardless of the signs of both the flux and the gradient of tracer, whether positive or negative.  
\Eq{eq:kzzcalc} allows one to calculate the vertical diffusivity that, in the context of a 1D model, would cause the same diffusive vertical flux as occurs via dynamical mixing in the 3D model (supposing such a 1D model had the same vertical gradient of horizontal-mean tracer as exists in the GCM). Using this approach, we hence derive an effective diffusive vertical mixing rate from an inherently non-diffusive GCM. In the following, we show that $\Kzz$ depends both on the circulation and the strength of the chemical source/sink (either the chemical timescale for our chemical relaxation scheme, or the particle size and hence settling velocity for the aerosol tracer scheme). 
\subsubsection{Chemical relaxation tracers}
\indent \Fig{fig:kzz_tauchem_drag_teq} shows the $\Kzz$ calculated from our models with chemical relaxation tracers, calculated at individual timesteps and then time-averaged over the last $100 \ \mathrm{days}$ of model time. The three columns in \Fig{fig:kzz_tauchem_drag_teq} can be viewed as different 2D slices through a single 3D parameter space of $\Kzz$ as a function of $T_\mathrm{eq}$, $\tau_\mathrm{drag}$, and $\tau_\mathrm{chem}$. We show results as a function of varying equilibrium temperature and drag timescale (for our models with fixed $\tau_\mathrm{chem} = 1.6 \times 10^5~\mathrm{s}$), varying chemical timescale and equilibrium temperature (for our models with drag only at the bottom of the domain), and varying chemical timescale and drag timescale (for our models with $T_{\mathrm{eq}} = 1500~\mathrm{K}$). As one might naively expect, $\Kzz$ generally increases with increasing equilibrium temperature and increasing drag timescale, with both of the trends due to faster winds. \\
\indent However, we find a local maximum in $\Kzz$ at intermediate drag timescales of $\sim 10^6~\mathrm{s}$ (\Fig{fig:kzz_tauchem_drag_teq}, left). In this case, the local maximum of tracer abundance from our GCM experiments at intermediate $\tau_\mathrm{drag}$ and high $T_\mathrm{eq}$ as shown in \Fig{fig:tauchem} manifests itself in \Fig{fig:kzz_tauchem_drag_teq} as an increased $\Kzz$. Note that the result of a local maximum in $\Kzz$ with increasing $\tau_\mathrm{drag}$ is independent of the assumed value of $\tau_\mathrm{chem}$, as it naturally results from the transition between day-to-night flow and superrotation discussed in \Sec{sec:vertvel}. That is, in the superrotating regime, the tracer is horizontally transported between upwelling and downwelling columns on short timescales. This results in a weakened spatial correlation between tracer anomalies and vertical velocities on an isobar, and therefore weakened net vertical transport compared to when the circulation is dominated by day-to-night flow. In the case of pure day-to-night flow, vertical transport is upward everywhere on the dayside and the tracer anomalies are better correlated to vertical velocities, enabling particles to be vertically mixed much more efficiently. \\
\indent Over much of parameter space, $\Kzz$ increases with increasing $\tau_\mathrm{chem}$, as species have more time to mix before they relax back to equilibrium. However, there are also are local minima and maxima in $\Kzz$ with increasing $\tau_\mathrm{chem}$, for example in the middle panels of \Fig{fig:kzz_tauchem_drag_teq} when $\tau_\mathrm{chem} \sim 5 \times 10^5 \ \mathrm{s}$ and the at a similar value of $\tau_\mathrm{chem}$ in the right-hand panels of \Fig{fig:kzz_tauchem_drag_teq}. A similar non-monotonic dependence in $\Kzz$ with chemical timescale was also found by \cite{Zhang:2017,Zhang:2018aa}. These local minima/maxima can lead to an order-of-magnitude change in $\Kzz$ with only a factor of $\sim 2$ change in $\tau_\mathrm{chem}$. Future work is required to understand the mechanism driving the location of local minima and maxima in $\Kzz$ with varying chemical relaxation timescale.
\subsubsection{Aerosol tracers}
\begin{figure*}[ht!]
	\centering
	\includegraphics[width=1\textwidth]{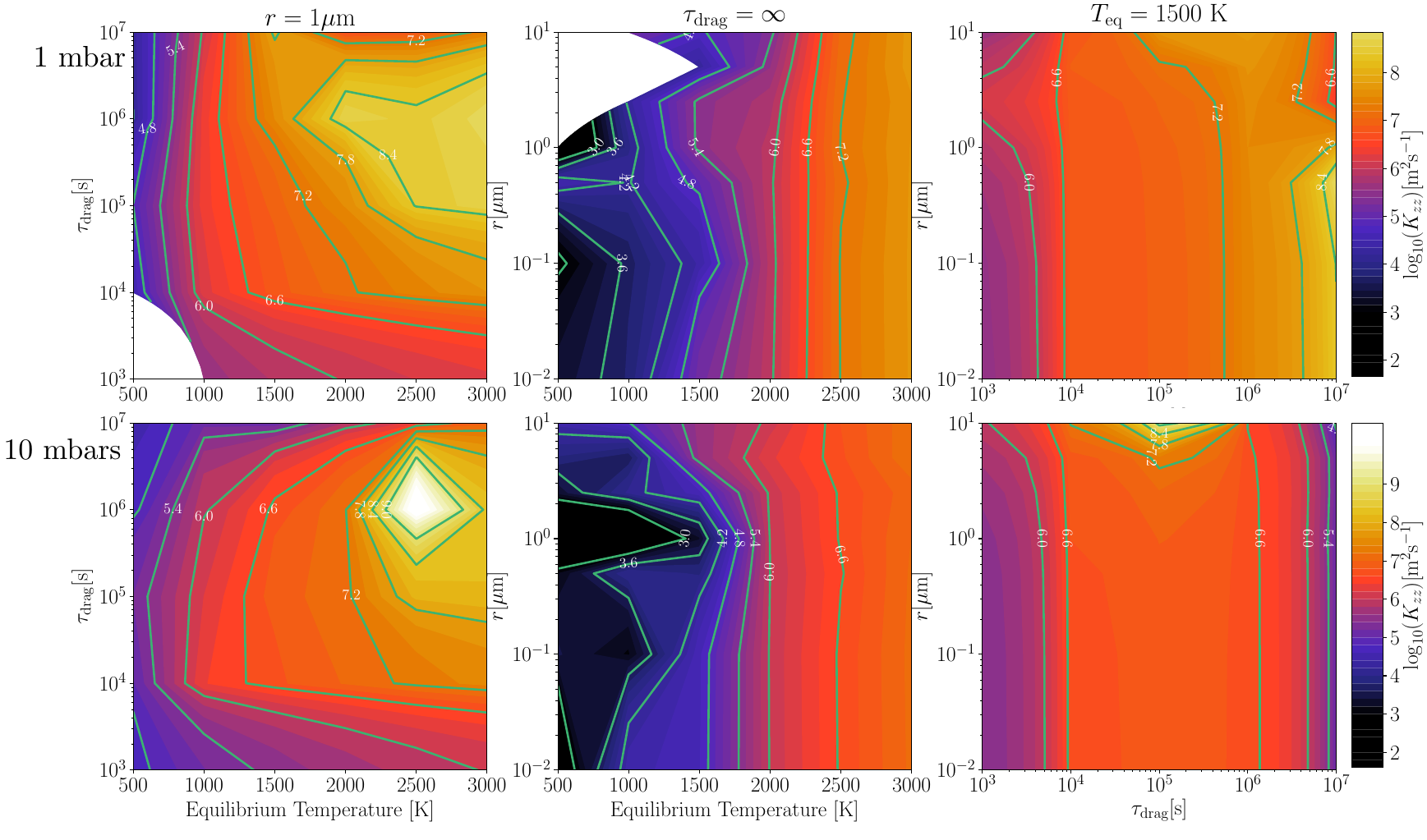}
	\caption{Vertical diffusivity $\Kzz$ at $1$ and $10 \ \mathrm{mbar}$ pressure calculated from our suite of GCM experiments using an aerosol tracer that settles vertically. The figure depicts three 2D cuts through a 3D parameter space of $\Kzz$ as a function of $T_\mathrm{eq}$, $\tau_\mathrm{drag}$, and particle size. Specifically, from left to right, the panels show $\Kzz$ as a function of drag timescale and equilibrium temperature (with fixed particle size of $1~\mu\mathrm{m}$), particle radius and equilibrium temperature (with fixed $\tau_{\mathrm{drag}} = \infty$), and particle size and drag timescale (with fixed $T_\mathrm{eq} = 1500 \ \mathrm{K}$). The top row of plots shows $\Kzz$ at a pressure of $1 \ \mathrm{mbar}$, and the bottom row of plots shows $\Kzz$ at $10 \ \mathrm{mbars}$. The top middle panel can be considered as a $\Kzz$ version of \Fig{fig:settling_1mbar}. Brighter colors correspond to larger $\Kzz$, and contours show $\mathrm{log}_{10}\Kzz$ in $\mathrm{m}^2\mathrm{s}^{-1}$. As we found for chemical relaxation tracers, with a fixed particle size (left panel) there is a maximum in the $\Kzz$ at intermediate drag timescales $\sim 10^6~\mathrm{s}$. As previously found by the GCM investigation of \cite{parmentier_2013}, $\Kzz$ is relatively weakly dependent on the particle size compared to other parameters (middle and right panels). Note that we cannot calculate a $\Kzz$ from our simulations with weak irradiation, strong drag, and/or large particle sizes because the settling timescale is very short relative to the dynamical timescale.}
	\label{fig:kzz_parsize_drag_teq}
\end{figure*}
We now turn to examine how $\Kzz$ varies with particle size, equilibrium temperature, and drag strength for our aerosol tracers that settle with a source/sink given by \Eq{eq:settletrac}. \Fig{fig:kzz_parsize_drag_teq} shows our numerical results for how $\Kzz$ depends on these parameters. Note that we cannot calculate a $K_\mathrm{zz}$ from our simulations with weak irradiation ($T_\mathrm{eq} = 500-1000 \ \mathrm{K}$) and large particles ($1-10 \ \mu\mathrm{m}$) at a pressure of $1 \ \mathrm{mbar}$ because the tracers settle out of the atmosphere extremely rapidly (the settling timescale is much shorter than the dynamical timescale). \\
\indent The left-hand panels of \Fig{fig:kzz_parsize_drag_teq} show how $\Kzz$ depends on drag timescale and equilibrium temperature from simulations with a fixed particle size of $1 \ \mu\mathrm{m}$. As in \Fig{fig:kzz_tauchem_drag_teq}, we find that there is a maximum in $\Kzz$ at intermediate $\tau_\mathrm{drag} \sim 10^6~\mathrm{s}$. This maximum occurs at the transition from superrotation for longer drag timescales to day-to-night flow with shorter drag timescales. This result implies that the increase in $\Kzz$ in the day-to-night flow regime does not depend on the tracer scheme considered, as it occurs for both our chemical relaxation and aerosol tracers. The reason is the same --- the vertical velocity and tracer variations (on isobars) tend to be well correlated in the day-night flow regime and less well correlated in the superrotating regime, promoting larger $\Kzz$ in the former case. And yet, if the drag is too strong in the day-night regime, the wind speeds and therefore mixing and $\Kzz$ become weak. Therefore, maximal $\Kzz$ occurs at the ``happy medium'' where drag is just strong enough to damp the equatorial jet but not stronger; in other words, right at the transition from superrotating to day-night flow as a function of $\tau_\mathrm{drag}$. \\
\indent The center panels of \Fig{fig:kzz_parsize_drag_teq} show the dependence of $\Kzz$ on particle size and equilibrium temperature. Similarly to \cite{parmentier_2013}, we find that $\Kzz$ is relatively insensitive to particle size, with variations nominally less than an order of magnitude in $\Kzz$ over a three order of magnitude range in particle radius. This is because, even though varying the particle size affects both the flux of tracer ($\langle \rho q w\rangle$) and its vertical gradient ($\langle \rho \partial q/\partial z\rangle$), it causes similar changes in their magnitude. Essentially, when the particle size is large, the efficiency of particle settling leads to a large vertical gradient in tracer abundance, and the vertical advection of this large gradient (upward in ascending regions and downward in descending regions) leads to large horizontal variations of tracer on isobars and therefore a large vertical tracer flux $\langle \rho q w\rangle$.  On the other hand, when the particle size is small, the tracers become more homogenized vertically, leading to a small vertical gradient in tracer abundance. The vertical advection of this weak background gradient leads to only small variations of tracer on an isobar, and therefore a small vertical tracer flux $\langle \rho q w\rangle$.  In this way, the interaction of dynamics with tracer mixing naturally causes the tracer flux and the vertical tracer gradient to respond in a similar way to variations in particle radius.  Since $\Kzz$ is the ratio of these quantities (Equation \ref{eq:kzzcalc}), this effect leads to a $\Kzz$ that does not depend strongly  on particle size, despite the fact that the tracer abundance itself {\it does} depend strongly on particle size (Figure \ref{fig:settling_1mbar}). This explanation helps to resolve the puzzle raised by \cite{parmentier_2013} that the derived $\Kzz$ was nearly independent of particle size in their GCM experiments. \\
\indent The right hand panels of \Fig{fig:kzz_parsize_drag_teq} show that there is a strong local maximum in $\Kzz$ with increasing particle size in the limit of weak drag, as is the case for our chemical relaxation tracers as shown in \Fig{fig:kzz_tauchem_drag_teq}. This local maximum occurs at a particle size of $0.5 \mu \mathrm{m}$. Particles with this size have a settling timescale (see Figure 2 of \citealp{parmentier_2013}) that is similar to the chemical relaxation timescale ($\sim 5 \times 10^5 \ \mathrm{s}$) at which we found local minima/maxima in \Fig{fig:kzz_tauchem_drag_teq}. Potentially, the same mechanism could be driving the local minima/maxima in $\Kzz$ with varying chemical relaxation and settling timescales. 
\section{A predictive theory for vertical mixing rates}
\label{sec:theory}
\subsection{Vertical wind speed}
\label{sec:vertwinds}
In hot Jupiter atmospheres, strong winds are driven by the large difference in received stellar flux between the dayside and nightside of the planet. In \cite{Komacek:2015}, we derived analytic expressions for the dayside-nightside temperature differences in hot Jupiter atmospheres, which in turn enables a prediction of the characteristic horizontal and vertical wind speeds. We presented these analytic solutions for wind speeds considering different possible force-balances, with the dayside-to-nightside pressure gradient (which is determined by the day-to-night temperature gradient) balanced by either the Coriolis force, advective terms, or frictional drag (see Equation 32 of \citealp{Komacek:2015}). \cite{Zhang:2016} improved on this theory by solving for the dayside-to-nightside temperature difference and characteristic horizontal wind speeds considering the combined effects of Coriolis, advection, and drag forces (see their Appendix A)\footnote{The theory in \cite{Zhang:2016} is essentially the same as in \cite{Komacek:2015}, in that the predictions for day-night temperature differences and wind speeds between the two studies are identical for any given force balance between pressure-gradient forces and either Coriolis, advective, or drag forces. However, \cite{Zhang:2016} provided a more convenient analytical representation of the transitions between the regimes such that all possible regimes can be incorporated into a single analytical expression, given (in the case of horizontal wind speed) by \Eq{eq:U}.}. \cite{Zhang:2016} also showed that the characteristic horizontal wind speeds predicted by this theory generally matches well the root-mean-square (RMS) horizontal wind speeds calculated from general circulation models (see their Figure 11). \\
\indent Here we utilize the solution of \cite{Zhang:2016} for horizontal wind speeds to estimate the characteristic global  vertical wind speeds in hot Jupiter atmospheres. Their solution for the characteristic horizontal wind speed is (see their Equation A.3)
\begin{equation}
\label{eq:U}
\mathcal{U} \sim \frac{2\gamma U_{\mathrm{eq}}}{\alpha + \sqrt{\alpha^2 + 4\gamma^2}} \mathrm{,}
\end{equation}
where, as in \cite{Komacek:2017},
\begin{equation}
\label{eq:alpha}
\alpha = 1 + \frac{\left(\Omega + \tau^{-1}_{\mathrm{drag}}\right)\tau^2_{\mathrm{wave}}}{\tau_{\mathrm{rad}} \Delta \mathrm{ln} p} \mathrm{,}
\end{equation}
and
\begin{equation}
\label{eq:gamma}
\gamma = \frac{\tau^2_{\mathrm{wave}}}{\tau_{\mathrm{rad}}\tau_{\mathrm{adv,eq}}\Delta \mathrm{ln} p} \mathrm{.}
\end{equation}
All variables in Equations (\ref{eq:U})-(\ref{eq:gamma}) have the same meaning as in \cite{Komacek:2017}\footnote{The characteristic horizontal wind speed is $\mathcal{U}$, the rotation rate is $\Omega$, the characteristic drag timescale is $\tau_{\mathrm{drag}}$, the (Kelvin) wave propagation timescale across a hemisphere is $\tau_{\mathrm{wave}} = a/(NH)$. Here, $a$ is the radius of the planet, $N$ is the Brunt-V{\"a}is{\"a}l{\"a} frequency, and $H=RT/g$ is the scale height, where $R$ is the specific gas constant, $T$ is temperature and $g$ is gravitational acceleration. The radiative timescale is $\tau_{\mathrm{rad}}$. $\Delta \mathrm{ln}p$ is the number of scale heights from the pressure of interest to the deep level at which the dayside-nightside temperature difference goes to zero, assumed as in \cite{Komacek:2017} to be $10 \ \mathrm{bars}$. $\tau_{\mathrm{adv,eq}} = a\sqrt{2/\left(R \Delta T_{\mathrm{eq}} \Delta \mathrm{ln}p\right)}$ is the advective timescale that a cyclostrophic wind induced by the day-night temperature difference in radiative equilibrium would have, where $\Delta T_{\mathrm{eq}}$ is the day-night temperature difference in radiative equilibrium. The speed of this maximum cyclostrophic wind is $U_{\mathrm{eq}} =  a/\tau_{\mathrm{adv,eq}}$.}. To relate the horizontal wind speed to the vertical wind speed, we utilize the scaled version of the continuity equation derived in \cite{Komacek:2015}:
\begin{equation}
\label{eq:uwscaling}
\frac{\mathcal{U}}{a} \sim \frac{\mathcal{W}}{H} \mathrm{.}
\end{equation}
Using \Eq{eq:uwscaling}, we find our final expression for vertical wind speeds:
\begin{equation}
\label{eq:wfinal}
\mathcal{W} \sim \frac{H}{a} \left(\frac{2\gamma U_{\mathrm{eq}}}{\alpha + \sqrt{\alpha^2 + 4\gamma^2}}\right) \mathrm{.}
\end{equation}
Note that the expression in \Eq{eq:wfinal} is always positive, and is intended to represent the characteristic magnitude of vertical wind speed as a function of height -- corresponding to upward on the dayside and downward on the nightside.  Importantly, our theory predicts that the vertical velocity is vertically coherent in the sense that, at a given location (whether day or night), it has the same sign vertically over many scale heights.
\\ \indent As an example, we use parameters of the typical hot Jupiter HD 209458b and assume no atmospheric drag, an equilibrium day-night temperature contrast equal to the equilibrium temperature of the planet ($\Delta T_\mathrm{eq} \sim T_\mathrm{eq}$), and a radiative timescale that scales as $\tau_\mathrm{rad} = 10^5~\mathrm{s}\left(p/100~\mathrm{mbar}\right)$ \citep{Komacek:2017}. Doing so, \Eq{eq:wfinal} predicts a characteristic vertical wind speed of $\sim 4 \ \mathrm{m} \ \mathrm{s}^{-1}$ at a pressure of $0.1 \ \mathrm{bars}$, rising to $\sim 20 \ \mathrm{m} \ \mathrm{s}^{-1}$ at $10 \ \mathrm{mbars}$. Note that our predicted wind speeds are also dependent on rotation rate -- if we change the rotation rate of the planet to be similar to that of WASP-43b (which, assuming that it is spin-synchronized, has a very short rotation period of $\approx 0.81~\mathrm{days}$), we predict a wind speed at $10~\mathrm{mbars}$ of $\sim 12\ \mathrm{m} \ \mathrm{s}^{-1}$, substantially smaller than the $\sim 20 \ \mathrm{m} \ \mathrm{s}^{-1}$ at $10 \ \mathrm{mbars}$ predicted for a planet with the rotation period of HD 209458b ($\approx 3.5~\mathrm{days}$). In general, our theory predicts that vertical velocities will increase with increasing incident stellar flux but decrease with increasing rotation rate, surface gravity, and frictional drag strength.
\subsection{Predicting $K_{\mathrm{zz}}$}
\label{sec:kzzpredict}
Though the theory of \cite{Komacek:2015} directly predicts the vertical wind speeds, this must be augmented in order to estimate an effective vertical diffusivity (i.e. $K_{\mathrm{zz}}$) similar to those used in one-dimensional chemical models of hot Jupiter atmospheres. Many previous models of hot Jupiter atmospheres have 
used mixing-length theory to estimate $\Kzz$ \citep{Marley:2014}, where $K_\mathrm{zz} \sim wl$ is an effective turbulent diffusivity that is related to the mixing length $l$ (assumed to be of order the scale height) and vertical wind speed $w$. Given a (convective) heat flux, one can then estimate the vertical wind speed and hence $K_\mathrm{zz}$. However, such mixing-length theory scalings only apply in the convective zone, while observable regions of hot Jupiter atmospheres are well above the radiative-convective boundary (RCB). The RCB can be as deep as thousands of bars, though it is potentially much shallower ($\sim 10-100 \ \mathrm{bars}$) on the cooler nightside and poles \citep{rauscher_showman_2013}. As a result, standard mixing-length theory is not an appropriate way to model the circulation in the observable parts of the atmospheres of hot Jupiters, as they are not convectively mixed. \\
\indent There has been no prior analytic theory for mixing and hence $\Kzz$ in the stably stratified regions of hot Jupiters. In this section, we seek such a self-consistent predictive theory for the specific case of our chemical relaxation tracers, whose source/sink is more amenable to an analytical treatment. Previous estimates of $\Kzz$ in hot Jupiter atmospheres have relied upon calculating $\Kzz$ from general circulation models \citep{Cooper:2006,Lewis:2010,Heng:2011a,Moses:2011,Venot:2012}, often assuming that $K_{\mathrm{zz}}$ is the product of a vertical velocity and vertical length scale, similar to mixing-length theory. However, \cite{parmentier_2013} showed that this approach greatly overestimates the more realistic calculation of mixing rates through including passive tracers in a GCM. Moreover, this is simply a numerical estimate that can only be performed with a numerical simulation of the flow, and such a calculation provides no fundamental understanding of how the mixing (and $\Kzz$) should scale with planetary parameters.  \\
\indent In this section, we aim to obtain an analytical solution for $\Kzz$ that is relevant to the radiative zone of extrasolar giant planets. As a result, here we apply a modified version of the transport parameterization developed by \cite{Holton:1986} for Earth's stratosphere (which like hot Jupiter atmospheres is stably stratified) to hot Jupiters in order to calculate analytic estimates for $K_\mathrm{zz}$ in a dynamically consistent fashion. This theory is similar to that developed by \cite{Zhang:2017} for short-lived tracers in a diffusive regime. This theoretical approach assumes a Newtonian relaxation source/sink, relevant to the chemical relaxation tracers in our GCM. To this end, in the following we confine our theory-GCM comparisons to those GCM simulations that adopt the same type of chemical source/sink forcing, and do not compare our theory to the tracers with cloud settling forcing.
\subsubsection{General formalism}
We consider a system where passive tracers are advected by the flow but do not influence the dynamics. The tracers will be mixed vertically by atmospheric circulation if there is a local correlation between tracer abundance and vertical velocity. For example, if tracer abundances are high where the vertical velocities are upward, then tracer is mixed upward (and the reverse is also true). In sum, vertical  transport depends strongly on the horizontal distribution of tracer and the correlation between this distribution and the vertical velocity field.  Note that if there is a vertical gradient of the mean tracer abundance, vertical motions will produce tracer perturbations on isobars that are correlated with the vertical velocity field. However, horizontal mixing and chemical losses can act to damp the horizontal tracer perturbations on isobars. As a result, the amplitude of tracer variations on isobars, and hence the vertical flux of tracer transported by the circulation, will depend on a balance between vertical advection, horizontal mixing, and chemical loss. \\
\indent This qualitative understanding was quantified by \cite{Holton:1986} for the case of a global-scale meridional overturning circulation. Here we modify this model in order to apply it to study the two-dimensional circulation (in height and longitude) in the equatorial regions of a spin-synchronized gas giant. In this specific case, the circulation at large scales should dominate mixing, since the circulation is driven by the extreme day-to-night irradiation difference \citep{showman_2002,perna_2012,parmentier_2013,Komacek:2015}. \\
\indent First, let $X$ be the mixing ratio of tracer. We can decompose $X$ into a Fourier series as
\begin{equation}
X(x,z^{\star}) = \sum^{\infty}_{n=0} X_n(z^{\star})\mathrm{cos}\left(\frac{nx}{a}\right) \mathrm{,}
\end{equation}
where $x$ is the eastward distance, $z^{\star} \equiv - \mathrm{ln}(p/p_{0})$ is the log-pressure coordinate with $p_{0}$ a reference pressure, $a$ is the planetary radius, and $n$ is the dimensionless planetary zonal wavenumber, that is, the integer number of wavelengths that fit around the circumference of the planet.  As in \cite{Holton:1986}, the time-evolution of tracer abundance is
\begin{equation}
\label{eq:dxdt}
\frac{dX}{dt} = S \mathrm{,}
\end{equation}
where $d/dt = \partial/\partial t + u \partial/\partial x + w^{\star}\partial/\partial z^{\star}$ is the advective derivative in log-pressure coordinates where $u$ is the horizontal velocity, $w^{\star} \equiv dz^{\star}/dt$ is the vertical velocity in units of scale heights per second, and $S$ represents the net source/sink of tracer. We further decompose the tracer abundance and source into a reference state that only depends on height and time, along with deviations from that reference state as 
\begin{equation}
\label{eq:Xdecomp}
X(x,z^{\star},t) = X_0(z^{\star},t) + \chi(x,z^{\star},t) \mathrm{,}
\end{equation}
\begin{equation}
\label{eq:Sdecomp}
S(x,z^{\star},t) = S_0(z^{\star},t) + s(x,z^{\star},t) \mathrm{.}
\end{equation}
Note that the deviations from the reference state are defined such that their zonal averages are zero, i.e. $\overline{\chi} = 0$ and $\overline{s} =0$, where the overbar denotes a zonal average. \\
\indent Inserting our decompositions (Equations \ref{eq:Xdecomp} and \ref{eq:Sdecomp}) into \Eq{eq:dxdt}, we find
\begin{equation}
\label{eq:tracerinit}
\frac{\partial(X_0 + \chi)}{\partial t} + u \frac{\partial \chi}{\partial x} + w^{\star} \frac{\partial(X_0 + \chi)}{\partial z^{\star}} = S_0 + s \mathrm{.} 
\end{equation}
The continuity equation in the two-dimensional primitive equations in log-pressure coordinates is \citep{Andrews:1987}
\begin{equation}
\label{eq:cont}
\frac{\partial u}{\partial x} + e^{z^{\star}} \frac{\partial\left(e^{-z^{\star}}w^{\star}\right)}{\partial z^{\star}} = 0 \mathrm{.}
\end{equation}
Note that since $\chi$ times \Eq{eq:cont} simply equals zero we can add their product to \Eq{eq:tracerinit} to find
\begin{equation}
\label{eq:tracer2}
\begin{aligned}
\frac{\partial(X_0 + \chi)}{\partial t} + & u \frac{\partial \chi}{\partial x} + w^{\star} \frac{\partial(X_0 + \chi)}{\partial z^{\star}} + \chi \frac{\partial u}{\partial x} + \chi e^{z^{\star}} \frac{\partial\left(e^{-z^{\star}}w^{\star}\right)}{\partial z^{\star}} \\ & = S_0 + s \mathrm{.} 
\end{aligned}
\end{equation}
One can use the product rule to show that
\begin{equation}
\chi e^{z^{\star}} \frac{\partial\left(e^{-z^{\star}}w^{\star}\right)}{\partial z^{\star}} = e^{z^{\star}} \frac{\partial\left(e^{-z^{\star}}w^{\star}\chi\right)}{\partial z^{\star}} - w^{\star} \frac{\partial \chi}{\partial z^{\star}} \mathrm{,}
\end{equation}
which when inserted into \Eq{eq:tracer2} yields
\begin{equation}
\label{eq:tracer3}
\frac{\partial(X_0 + \chi)}{\partial t} + \frac{\partial(u\chi)}{\partial x} + w^{\star} \frac{\partial X_0}{\partial z^{\star}} + e^{z^{\star}} \frac{\partial\left(e^{-z^{\star}}w^{\star}\chi\right)}{\partial z^{\star}} = S_0 + s \mathrm{.}
\end{equation}
We can now manipulate this expression to find equations that govern the tracer abundance. First, we zonally average \Eq{eq:tracer3}. Note that because $X_0$ is independent of $x$, the quantity $\overline{w^{\star}\partial X_0/\partial z^{\star}} = \overline{w^{\star}}\partial X_0/\partial z^{\star}$, where $\overline{w^{\star}} = 0$ from continuity (Equation \ref{eq:cont}). Additionally, $\overline{\partial (u\chi)/\partial x} = 0$ since the solution is longitudinally periodic. Through performing this zonal average, we find
\begin{equation}
\label{eq:zonalaverage}
\frac{\partial X_0}{\partial t} + e^{z^{\star}} \frac{\partial \left(e^{-z^{\star}}\overline{w^{\star}\chi}\right)}{\partial z^{\star}} = S_0 \mathrm{.}
\end{equation}
Lastly, we can subtract \Eq{eq:zonalaverage} from \Eq{eq:tracer3} to find
\begin{equation}
\label{eq:tracerfin}
\frac{\partial \chi}{\partial t} + \frac{\partial \left(u\chi\right)}{\partial x} + w^{\star} \frac{\partial X_0}{\partial z^{\star}} + e^{z^{\star}}\frac{\partial \left[e^{-z^{\star}}\left(w^{\star}\chi - \overline{w^{\star}\chi}\right)\right]}{\partial z^{\star}} = s \mathrm{.}
\end{equation}
\subsubsection{Diffusion equation and effective $\Kzz$} 
We now use assumptions similar to those of \cite{Holton:1986} and \cite{Zhang:2017} in order to translate \Eq{eq:tracerfin} into a diffusion equation for the mean tracer abundance $X_0$. First, we neglect the term $e^{z^{\star}}\partial \left[e^{-z^{\star}}\left(w^{\star}\chi - \overline{w^{\star}\chi}\right)\right]/\partial z^{\star}$, as it corresponds to the vertical advection of the tracer perturbation $\chi$ by the flow and we assume that this is small compared to the vertical advection of $X_0$ by the flow, $w^{\star} \partial X_0/\partial z^{\star}$. This approximation should be valid, as we can estimate that $e^{z^{\star}}\partial \left[e^{-z^{\star}}\left(w^{\star}\chi - \overline{w^{\star}\chi}\right)\right]/\partial z^{\star}$ scales as the larger of $w^{\star}\Delta \chi/H$ and $\chi \Delta w^{\star}/H$, where $\Delta$ represents the characteristic variation over a scale height. Comparing this to the magnitude of $w^{\star} \partial X_0/\partial z^{\star}$, which is $w^{\star} \Delta X_0/H$, and noting that $\Delta w^{\star} \sim w^{\star}$, we find that neglect of this term is valid if $\chi \ll \Delta X_0$. \\
\indent Secondly, we assume that the horizontal gradient of the horizontal eddy flux term $\partial(u\chi)/\partial x$ can be approximated as a horizontal eddy diffusion term, $-D \partial^2\chi/\partial x^2$, where $D$ is the horizontal diffusivity. Applying these two assumptions, \Eq{eq:tracerfin} becomes
\begin{equation}
\label{eq:difffirst}
\frac{\partial \chi}{\partial t} +  w^{\star} \frac{\partial X_0}{\partial z^{\star}} = D \frac{\partial^2 \chi}{\partial x^2} + s \mathrm{,}
\end{equation}
which is equivalent to Equation (7) of \cite{Holton:1986}. To further manipulate \Eq{eq:difffirst}, we assume that (as might be expected for an overturning circulation) the vertical velocity has a sinusoidal dependence, i.e. $w^{\star}(x,z^{\star}) = \hat{w}^{\star}(z^{\star}) e^{inx/a}$, where $\hat{w}^{\star}$ is the amplitude of the assumed sinusoidal variation of $w^{\star}$ with longitude. Then we need to only consider the projection of tracer onto this mode, that is, $\chi(x,z^{\star}) = \hat{\chi}(z^{\star}) e^{inx/a}$, where $\hat{\chi}$, which is a function of height, is the amplitude of the variation of $\chi$ with longitude. Additionally, as in our GCM we parameterize the source/sink of tracer as $s = -\chi/\tau_\mathrm{chem}$. Inserting these into \Eq{eq:difffirst} and simplifying, we find
\begin{equation}
\label{eq:diffalmost}
\frac{\partial \hat{\chi}}{\partial t} + \hat{w}^{\star}\frac{\partial X_0}{\partial z^{\star}} = \frac{-D}{a^2}\hat{\chi} - \frac{\hat{\chi}}{\tau_\mathrm{chem}} = -\hat{\chi} \left(\frac{1}{\tau_\mathrm{chem}} + \frac{1}{\tau_\mathrm{diff}}\right) \mathrm{,}
\end{equation}
where we have defined a horizontal diffusion time $\tau_{\mathrm{diff}} = a^2/D$. \Eq{eq:diffalmost} is the hot Jupiter equivalent to Equation (11) of \cite{Holton:1986}. \\
\indent Now, assuming steady state ($\partial \hat{\chi}/\partial t \rightarrow 0$), we can solve for $\hat{\chi}$:
\begin{equation}
\label{eq:chi}
\hat{\chi} = - \frac{w^{\star}\frac{\partial X_0}{\partial z^{\star}}}{\left(\frac{1}{\tau_\mathrm{chem}} + \frac{1}{\tau_\mathrm{diff}}\right)} \mathrm{.}
\end{equation}
Plugging \Eq{eq:chi} into the zonal-mean tracer evolution expression (Equation \ref{eq:zonalaverage}) yields our desired expression (analogous to Equation 13 of \citealp{Holton:1986}), a diffusion equation for $X_0$:
\begin{equation}
\label{eq:findiff}
\frac{\partial X_0}{\partial t} - e^{z^{\star}} \frac{\partial \left(e^{-z^{\star}}\hat{K}_{\mathrm{zz}}\frac{\partial X_0}{\partial z^{\star}}\right)}{\partial z^{\star}} = S_0 \mathrm{.}
\end{equation}
In \Eq{eq:findiff}, 
\begin{equation}
\label{eq:kzzfull}
\hat{K}_{\mathrm{zz}} = \frac{(\hat{w}^{\star})^2}{2\left(\frac{1}{\tau_\mathrm{chem}} + \frac{1}{\tau_\mathrm{diff}}\right)} 
\end{equation}
is the vertical diffusivity in log-pressure coordinates, which has units of scale heights squared per second. \\
\indent  Under our assumptions, we find that the large-scale vertical motions in hot Jupiter atmospheres mix the tracer in a diffusive manner. The effective diffusivity of these motions depends on the vertical velocity, the chemical loss timescale, and the horizontal diffusion timescale. It is clear from \Eq{eq:kzzfull} that faster vertical velocities should increase the eddy diffusivity while faster chemical loss and faster horizontal mixing timescales decrease the eddy diffusivity. Note that in the limit of infinitely fast horizontal mixing and/or chemical loss, the tracer would be constant on isobars and hence $\Kzz$ would be zero regardless of the vertical velocities. As a result, if mixing in the horizontal direction is the dominant process (with a short $\tau_\mathrm{diff}$), we predict that $\Kzz$ will be small. A key point to stress (see also \citealp{Zhang:2017}) is that $\Kzz$ will be different for different chemical species (which have different values of $\tau_{\mathrm{chem}}$), even within a single atmosphere. Therefore, in general it is not correct to use a single $\Kzz$ profile for all chemical species in a 1D chemical model. \\
\indent To relate the expression for $\Kzz$ from \Eq{eq:kzzfull} to our analytically predicted vertical velocities from \Eq{eq:wfinal}, we utilize our analytic theory for the RMS vertical velocity. To zeroth order, ignoring the effects of small-scale turbulent mixing, the large-scale horizontal diffusivity is $D \sim a\mathcal{U}$. Using \Eq{eq:uwscaling} to relate the vertical and horizontal wind speeds, we can estimate $\tau_{\mathrm{diff}} \sim H/\mathcal{W}$. Additionally, note that we can relate $\hat{w}^{\star}H \approx \mathcal{W}$. Plugging these scalings into \Eq{eq:kzzfull}, we can estimate $\Kzz$:
\begin{equation}
\label{eq:kzz}
K_\mathrm{zz} \sim \frac{\mathcal{W}^2}{(\tau^{-1}_{\mathrm{chem}} + \frac{\mathcal{W}}{H})} \mathrm{.}
\end{equation}
Here $\Kzz$ has absorbed a factor of $H^2$ from the relationship between $\hat{w}^{\star}$ and $\mathcal{W}$ and now has its standard dimensions of length$^2$ time$^{-1}$. \Eq{eq:kzz} is equivalent to Equation (13) of \cite{Zhang:2017}. 
Note that in the limit of infinitely long chemical timescales, \Eq{eq:kzz} simplifies to the basic result predicted by mixing-length theory, $\Kzz \sim H \mathcal{W}$. However, if chemical timescales are comparable to or shorter than the dynamical timescale $H/\mathcal{W}$, $K_{\mathrm{zz}}$ requires knowledge of the chemical timescale to calculate. Next, we compare the predictions of our theory to the GCM results that were presented in \Sec{sec:numerics} in order to test our predictions for the vertical velocity and vertical mixing rates.
\subsection{Comparison between analytic theory and general circulation models}
\label{sec:theorymodelcomp}
\subsubsection{Vertical velocities}
\begin{figure*}
	\centering
	\includegraphics[width=0.92\textwidth]{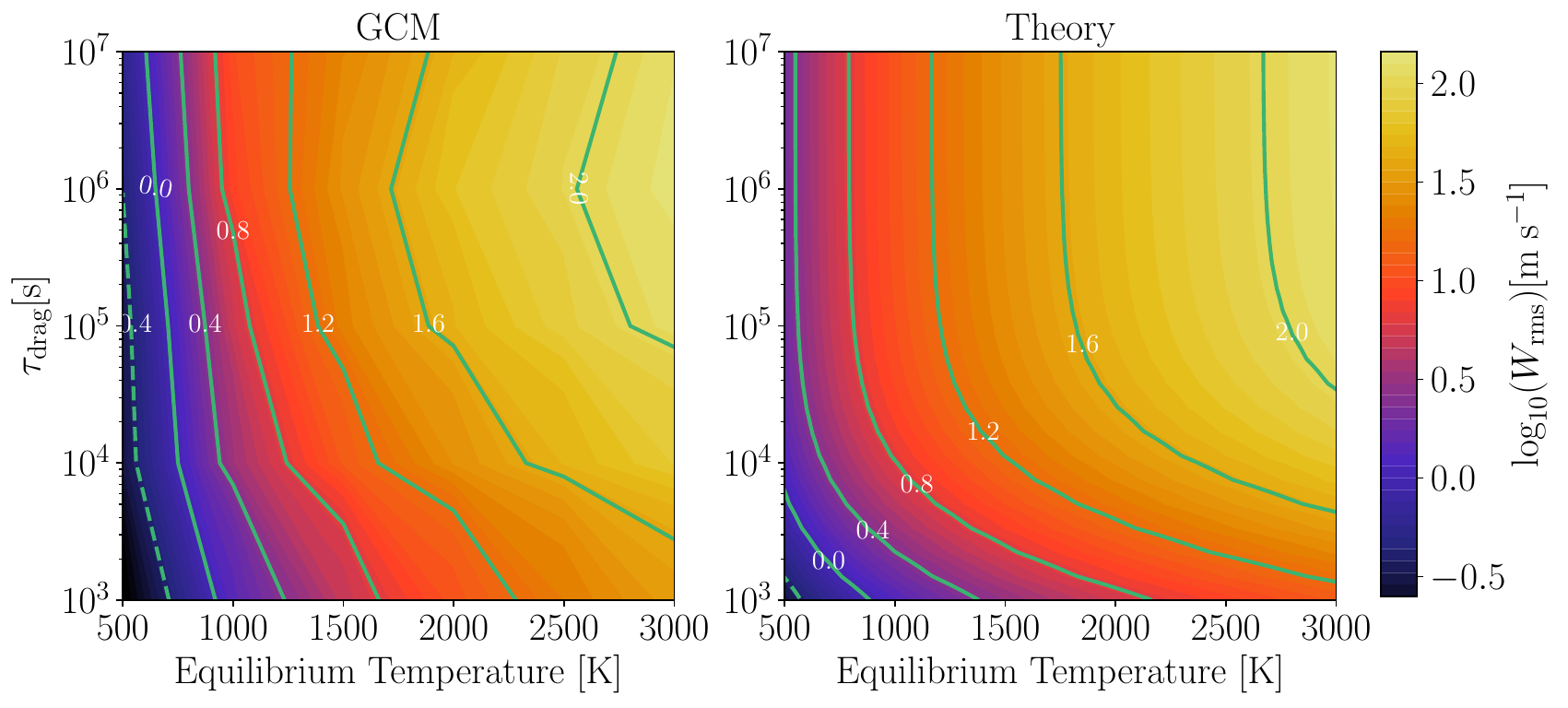}
	\caption{Theoretical predictions for the characteristic vertical wind speeds from \Eq{eq:wfinal} as a function of drag timescale and equilibrium temperature compared to RMS vertical velocity from our GCM experiments presented in \Sec{sec:numerics}. Brighter colors correspond to larger vertical wind speeds, and contours show $\mathrm{log}_{10}W$ in $\mathrm{m}\mathrm{s}^{-1}$. The dashed contour corresponds to a negative value of $\mathrm{log}_{10}W$. This comparison is done at a pressure of $1 \ \mathrm{mbar}$. As in \cite{Komacek:2015}, the theory matches well the general trends in vertical velocity, and correctly predicts the transition in drag timescale ($\sim 10^5 \ \mathrm{s}$) where vertical velocities no longer depend on the drag strength.  }
	\label{fig:Wcomp}
\end{figure*}
To establish the basic validity of the theory from \cite{Komacek:2015}, we first examine how our characteristic vertical wind speeds from \Eq{eq:wfinal} compare with the global root-mean-square (RMS) vertical velocities from our double-grey GCM. To do so, we utilize the same assumptions as in \cite{Komacek:2017}, that is:
\begin{enumerate}
\item $\Delta T_{\mathrm{eq}} \sim T_\mathrm{eq}$, as in radiative equilibrium the nightside should be extremely cold relative to the dayside.
\item We take the radiative timescale as a power-law $\tau_\mathrm{rad} \propto p T^{-3}_\mathrm{eq}$ as in \cite{showman_2002} and \cite{Ginzburg:2015a}, setting $\tau_{\mathrm{rad}} = 10^5 \ \mathrm{s}$ for $p = 100 \ \mathrm{mbar}$ and $T_\mathrm{eq} = 1800 \ \mathrm{K}$. 
\end{enumerate}
We calculate the RMS vertical velocities on isobars from our GCM experiments as in \cite{Komacek:2015},
\begin{equation}
W_{\mathrm{rms}}(p) = \sqrt{\frac{\int_0^{4\pi a^2} w^2 dA}{A}} \mathrm{,}
\end{equation}
where the integral is taken over the globe, with $A$ the horizontal area of the globe and $w$ the vertical velocity at a given pressure level. 
\\ \indent \Fig{fig:Wcomp} shows the comparison between our theoretical predictions and GCM results for how the vertical velocity varies with equilibrium temperature and drag timescale at a pressure of $1 \ \mathrm{mbar}$. The theory and GCM results show that vertical velocity increases with both increasing equilibrium temperature and increasing $\tau_\mathrm{drag}$. The theory broadly matches the magnitude of vertical velocities. 
The theory also correctly predicts the transition where drag no longer strongly influences the vertical velocity, which occurs between $\tau_{\mathrm{drag}} = 10^4 - 10^5 \ \mathrm{s}$ where drag becomes weaker than the Coriolis force. Note that the theory also predicts that the vertical velocity will decrease with increasing pressure, due to the reduced dayside-to-nightside forcing amplitude \citep{Komacek:2015,Zhang:2016}. In the next section, we will use our theoretical prediction of vertical velocities to estimate $\Kzz$ over a range of equilibrium temperature, drag strengths, rotation rates, and pressure levels. We do so because we are interested in a first-principles understanding of how $\Kzz$ depends on planetary parameters, rather than a calculation that requires detailed GCM simulations. The broad agreement between the scaling of vertical velocities with drag and equilibrium temperature predicted by our theory and calculated in our GCM gives us confidence in applying the formalism developed in \Sec{sec:kzzpredict} to predict $\Kzz$ using our analytic theory.
\subsubsection{Vertical mixing rates}
\label{sec:theorycompkzz}
\begin{figure*}
	\centering
	\includegraphics[width=0.9\textwidth]{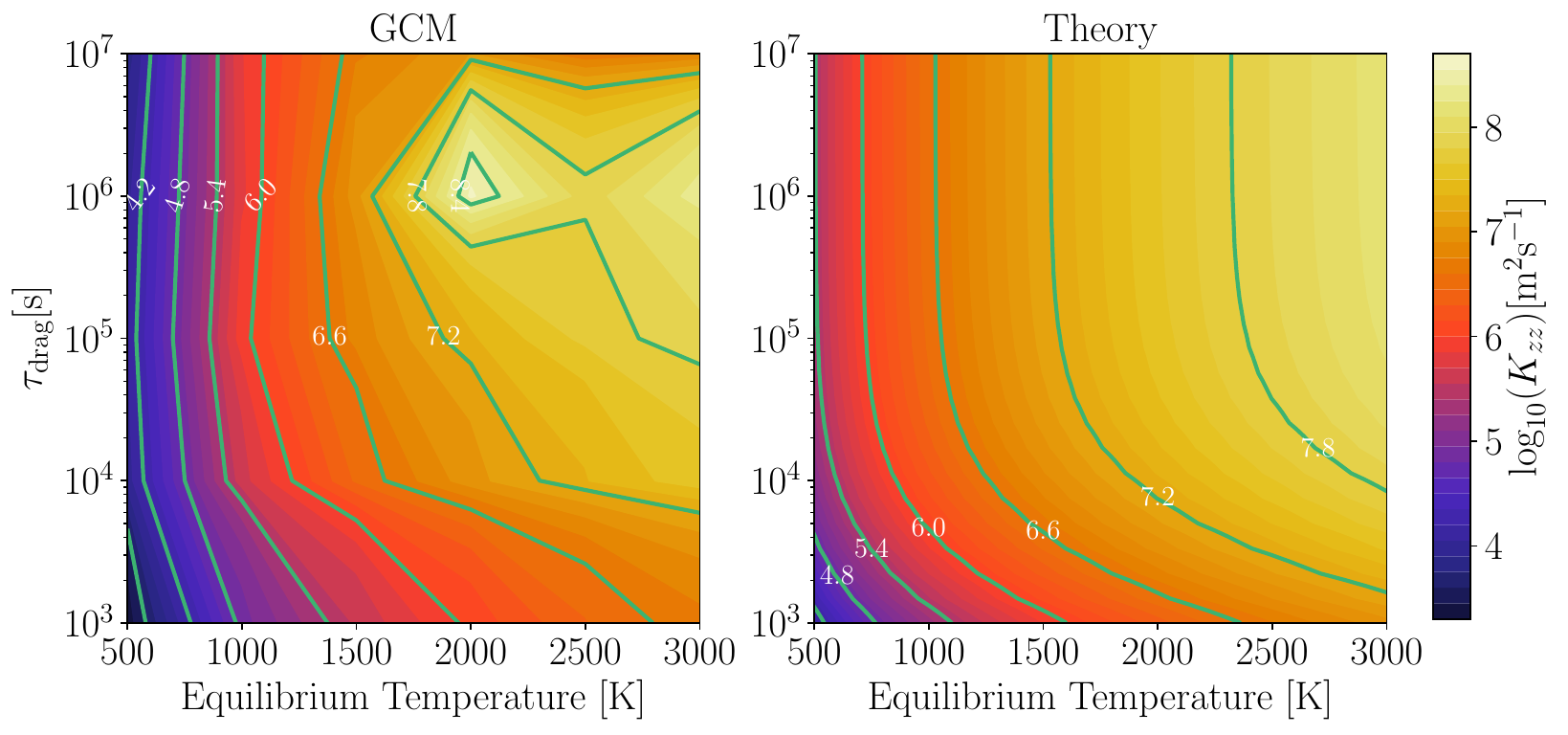}
	\caption{Vertical diffusivity $\Kzz$ from GCM experiments with $\tau_\mathrm{chem} = 1.6 \times 10^5 \ \mathrm{s}$ compared to theoretical predictions for $\Kzz$ as a function of drag timescale and equilibrium temperature. Brighter colors correspond to larger $\Kzz$, and contours show $\mathrm{log}_{10}\Kzz$ in $\mathrm{m}^2\mathrm{s}^{-1}$. This comparison is performed at a pressure of $1 \ \mathrm{mbar}$. The theory broadly captures the trends in $\Kzz$ calculated by our GCM, but does not encapsulate the reduction of $\Kzz$ at long drag timescales. This is because the flow becomes governed by a superrotating jet, while the theoretical predictions assume that the circulation is dominated by day-to-night flow. Additionally, note that the GCM predictions are qualitatively similar regardless of the chemical timescale or particle size assumed, as all have a decrease in $\Kzz$ with increasing $\tau_\mathrm{drag}$ from $10^6 \ \mathrm{s} - \infty$.}
	\label{fig:Kzzcomp}
\end{figure*}
\indent Using the vertical velocity from \Eq{eq:wfinal}, we can estimate $\Kzz$ using \Eq{eq:kzz}. \Fig{fig:Kzzcomp} shows a comparison between our theoretical predictions and GCM results for how $\Kzz$ varies with equilibrium temperature and drag timescale at a pressure of $1 \ \mathrm{mbar}$. Similar to the predicted vertical velocities, our analytic theory matches the general trends of increasing $\Kzz$ with increasing equilibrium temperature and drag strength. The theory also predicts to order-of-magnitude the quantitative value of $\Kzz$. However, it does not predict the local maximum in $\Kzz$ seen in our simulations. This is because the theory does not incorporate the superrotating jet itself. Instead, the theory simply assumes that all flow is from day-to-night, driven by the day-to-night temperature difference. As a result, as found in \cite{Zhang:2017,Zhang:2018aa} the theory over-predicts $\Kzz$ when drag is weak ($\tau_\mathrm{drag} \ge 10^7 \ \mathrm{s}$). This is also similar to the finding of \cite{parmentier_2013} that $\Kzz$ is over-predicted by a factor of $\sim 10-100$ by mixing length theory compared to simulations that do not include atmospheric drag. This is because in the weak-drag limit $\mathcal{W}/H \gg \tau^{-1}_\mathrm{chem}$ for any of our assumed $\tau_\mathrm{chem}$, and as a result \Eq{eq:kzz} simplifies to $\Kzz \sim \mathcal{W}H$, extremely similar to the relation predicted by mixing-length theory but with $\mathcal{W}$ found from first-principles.    \\ 
\begin{figure*}
	\centering
	\includegraphics[width=0.9\textwidth]{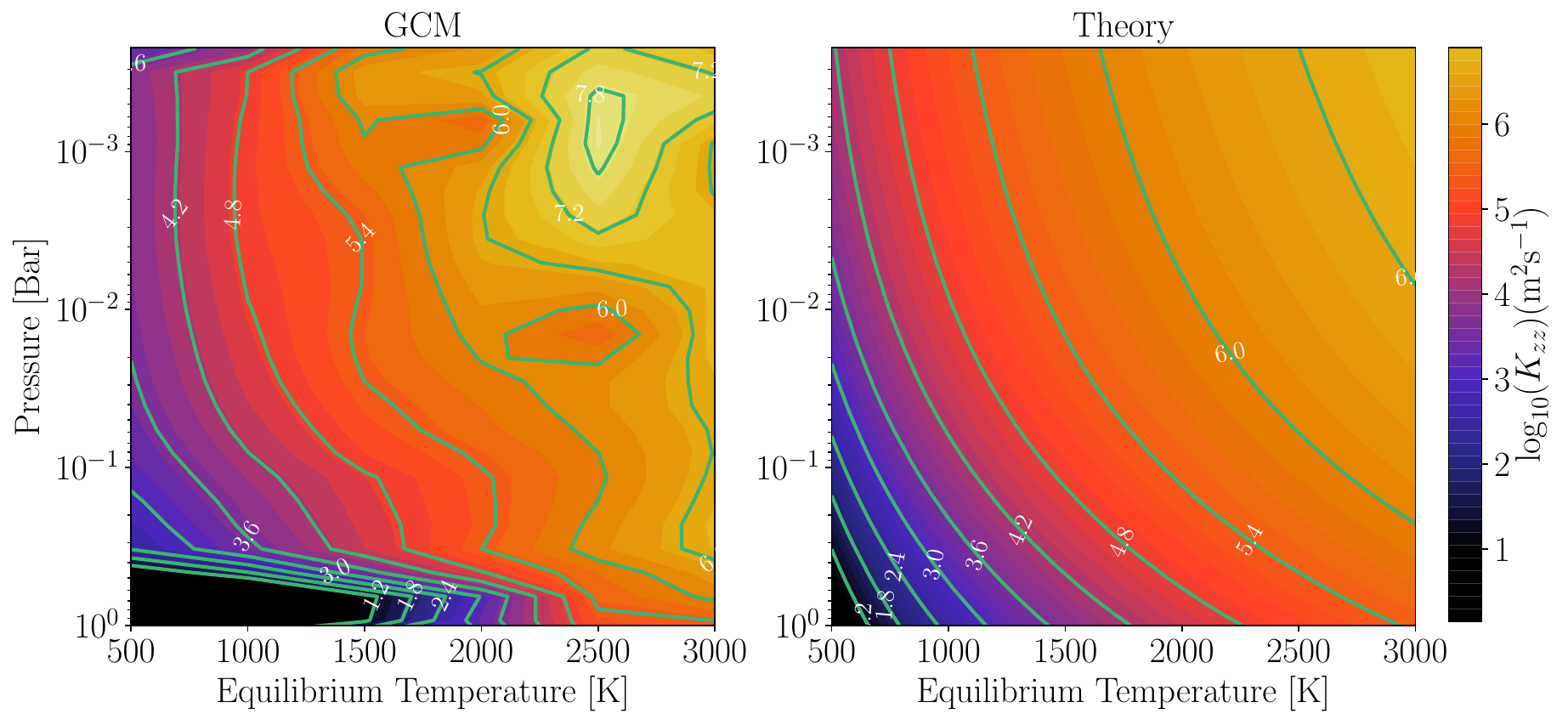}
	\caption{Vertical diffusivity $\Kzz$ from GCM experiments with $\tau_\mathrm{chem} = 1.6 \times 10^5 \ \mathrm{s}$ compared to theoretical predictions for $\Kzz$ as a function of pressure and equilibrium temperature. Brighter colors correspond to larger $\Kzz$, and contours show $\mathrm{log}_{10}\Kzz$ in $\mathrm{m}^2\mathrm{s}^{-1}$. This comparison is performed at a drag timescale $\tau_\mathrm{drag} = \infty$. The theory broadly captures the increase in $\Kzz$ with decreasing pressure and increasing equilibrium temperature found in our GCM simulations. However, note that high in the atmosphere ($p \sim 1 \ \mathrm{mbar}$) and at high equilibrium temperatures $T_\mathrm{eq} \gtrsim 2500 \ \mathrm{K}$ there is a large increase in $\Kzz$ in our GCM simulations, not captured in the analytic theory.}
	\label{fig:Kzzcomp_pressure}
\end{figure*}
\indent Our analytic theory predicts that the vertical wind speed should increase with decreasing pressure and increasing equilibrium temperature due to the increasing dayside-to-nightside forcing amplitude. Because $\Kzz$ scales with the vertical wind speed squared, we expect that $\Kzz$ should also increase with decreasing pressure and increasing equilibrium temperature. \Fig{fig:Kzzcomp_pressure} compares our theoretical predictions for $\Kzz$ with varying pressure and equilibrium temperature to our GCM experiments. In this comparison, we assume $\tau_\mathrm{drag} = \infty$ and $\tau_\mathrm{chem} = 1.6 \times 10^5 \ \mathrm{s}$. We find that the $\Kzz$ predicted by the GCM generally increases with increasing equilibrium temperature and decreasing pressure, in accordance with the analytic expectations. However, there is a marked increase in $\Kzz$ at high equilibrium temperatures $\gtrsim 2500 \ \mathrm{K}$ and low pressures $p \sim 1 \ \mathrm{mbar}$, not expected from our analytic theory. This is likely because our analytic theory does not incorporate changes in the structure of the atmospheric circulation with changing equilibrium temperature or pressure, instead simply assuming day-to-night flow throughout the atmosphere. \\
\begin{figure}
	\centering
	\includegraphics[width=0.5\textwidth]{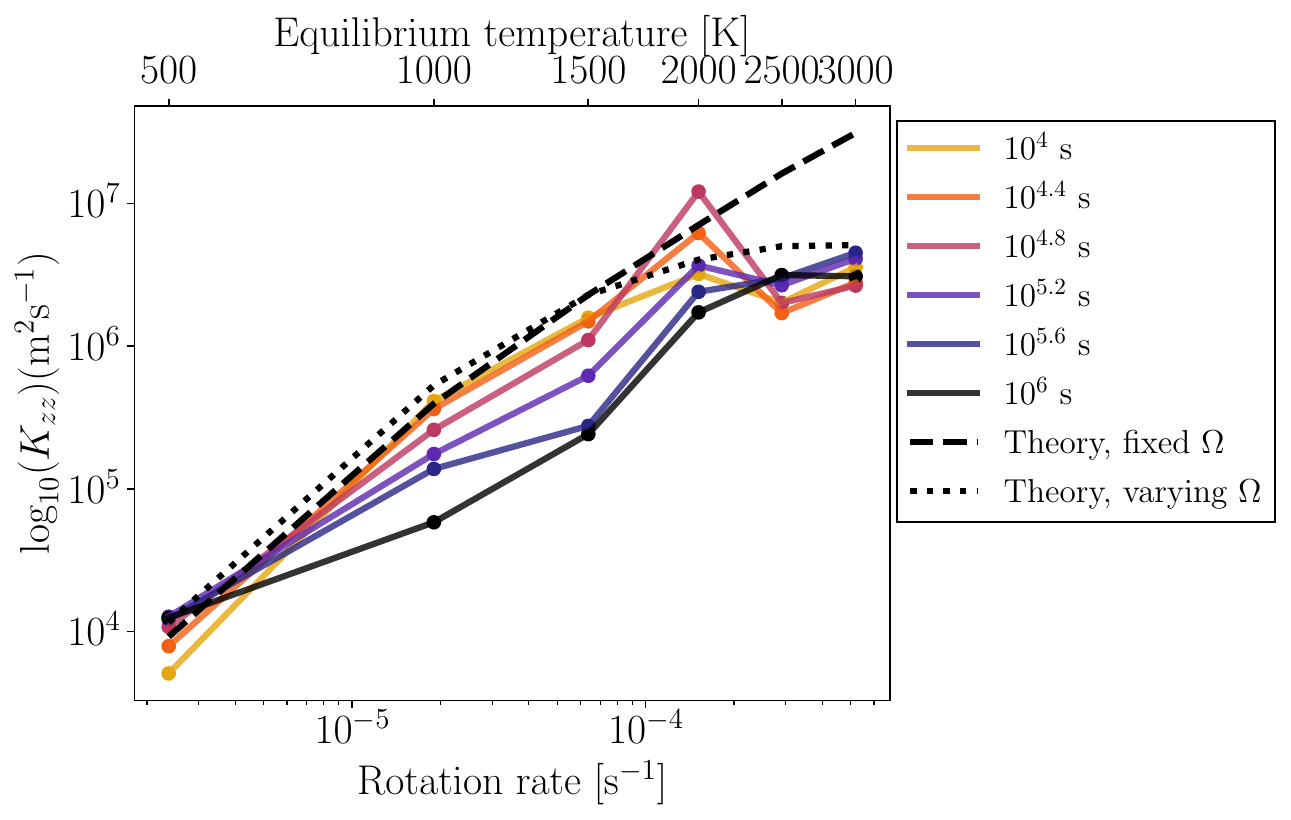}
	\caption{Vertical diffusivity $\Kzz$ from GCM experiments with consistently varying rotation rate and equilibrium temperature compared to our theoretical scaling. The solid lines show results from our GCM experiments at a pressure of $1 \ \mathrm{mbar}$, with dots representing individual simulations. We vary the equilibrium temperature from 500 to 3000 K for a planet around a star with properties of HD 189733, leading to rotation rates varying from $2.37 \times 10^{-6} \ \mathrm{s}^{-1}$ to $5.19 \times 10^{-4} \ \mathrm{s}^{-1}$. We show results for $\Kzz$ with varying chemical relaxation tracer timescale from $10^{4} - 10^6 \ \mathrm{s}$, with darker colors representing longer timescales. We overplot a modified theoretical scaling for $\Kzz$ assuming $\tau_\mathrm{chem} = 10^{5.2} \ \mathrm{s}$ both ignoring (dashed line) and including (dotted line) the effects of rotation on the vertical mixing rate. As predicted by our analytic theory, the slope of $\Kzz$ with rotation rate appears to flatten for rotation periods $\lesssim 1 \ \mathrm{day}$ (corresponding to rotation rates $\gtrsim 6 \times 10^{-5} \ \mathrm{s}$).}
	\label{fig:Kzzcomp_rotvarying}
\end{figure}
\indent We also analyze our predictions for vertical mixing rate with consistently varying rotation rate and equilibrium temperature. To do so, we use a suite of simulations with varying equilibrium temperature from 500-3000 K for a planet orbiting a host star fixed to have the properties of HD 189733 (see Section 3.5 of \citealp{Komacek:2017} for more details). The simulations have no applied drag at pressures below 10 bar, with $\tau_\mathrm{drag} = \infty$. We include the effects of consistently increasing rotation rate with equilibrium temperature according to Kepler's laws, leading to rotation rates that increase from $2.37 \times 10^{-6} \ \mathrm{s}^{-1}$ at $T_\mathrm{eq} = 500 \ \mathrm{K}$ to $5.19 \times 10^{-4} \ \mathrm{s}^{-1}$ at $T_\mathrm{eq} = 3000 \ \mathrm{K}$. \Fig{fig:Kzzcomp_rotvarying} shows our results for $\Kzz$ with varying chemical timescale from this additional suite of simulations. We find that $\Kzz$ increases with increasing equilibrium temperature and rotation rate, largely due to the strong effect of increasing incident stellar flux leading to more vigorous atmospheric circulation. We also plot our theoretical predictions for $\Kzz$ with fixed $\tau_\mathrm{chem} = 10^{5.2} \ \mathrm{s}$ solely including the effect of varying equilibrium temperature (dashed line, using a fixed rotation rate equal to that of the $T_\mathrm{eq} = 1500 \ \mathrm{K}$ simulation) and including both the effects of varying rotation rate and equilibrium temperature (dotted line). 
We find that the theory over-predicts $\Kzz$ at intermediate and fast rotation rates, similar to the over-prediction of $\Kzz$ with weak drag in \Fig{fig:Kzzcomp}. However, our theory including rotation (dotted lines in \Fig{fig:Kzzcomp_rotvarying}) does correctly predict that the slope of $\Kzz$ with rotation rate flattens for fast rotation rates $\gtrsim 6 \times 10^{-5} \ \mathrm{s}$ (corresponding to rotation periods $\lesssim 1 \ \mathrm{day}$).  \\
\indent In principle, it could be possible to incorporate the superrotating jet in a theoretical estimate for $\Kzz$. To do so, one must develop an understanding for why the correlation between vertical velocities and tracer abundance becomes weak in the superrotating regime relative to the day-night flow regime. Presumably there is a correlation factor between vertical velocities and tracer abundance that is high in the day-night regime and small in the superrotating regime. Understanding quantitatively how this correlation factor changes with planetary parameters is necessary to understand vertical transport in the superrotating regime in greater detail. Additionally, it would be useful to have a theory for the equilibrated wind speed of the superrotating jet, which does not currently exist (see the discussion in \citealp{Komacek:2017}). We stress that our analytic theory predicts the broad trends in $\Kzz$ with varying planetary parameters and hence has utility for understanding vertical transport in hot Jupiter atmospheres. 
\section{Discussion}
\label{sec:disc}
\subsection{Application of theory to one-dimensional models of spin-synchronized gas giants}
\subsubsection{Estimating vertical mixing rates}
\label{sec:cook}
The theory outlined in \Sec{sec:theory} predicts how vertical mixing rates in the atmospheres of spin-synchronized gas giants should vary with incident stellar flux, rotation rate, atmospheric pressure level, and potential frictional drag strength. As a result, this theory provides scaling relations for how vertical mixing rates should vary with planetary parameters. Our theory can be utilized to understand the vertical quenching of species in hot Jupiter atmospheres, including CO and CH$_4$. Our prediction for $\Kzz$ from \Eq{eq:kzz} from can be applied to 1D chemical models to constrain the quench level, which occurs when the dynamical timescale $\tau_\mathrm{dyn} \sim H^2/\Kzz$ is shorter than the chemical timescale. One can use our theory to estimate $\Kzz$ by the following method. First, estimate the global-mean vertical wind speed for given planetary parameters (e.g., radius, equilibrium temperature, surface gravity, rotation rate) using \Eq{eq:wfinal}. Then, calculate the relevant chemical relaxation timescale for the species of interest. Lastly, plug in the estimated vertical wind speed, chemical timescale, and scale height into \Eq{eq:kzz} to estimate the global $\Kzz$. Note that one could use the global-average vertical wind speed from a GCM calculation of a given planet rather than our theoretical prediction from \Eq{eq:wfinal} for a potentially more accurate prediction of $\Kzz$. \\
\indent When the dynamical timescale and chemical timescale are comparable, our prediction for $\Kzz$ depends on the chemical timescale itself. As a result, our prediction for $\Kzz$ depends on the chemical species considered, along with planetary parameters that determine the atmospheric wind speeds and scale height. The chemical timescale should change by many orders of magnitude with varying temperature \citep{Cooper:2006}, and as shown by \cite{Zhang:2017,Zhang:2018aa} may greatly vary between different chemical species. As a result, we stress that calculating $\Kzz$ requires knowledge of the specific chemical reaction being considered, which must be taken into account when using this theory to predict how disequilibrium chemical processes should be affected by the incident stellar flux, rotation rate, and atmospheric frictional drag. 
\subsubsection{Limitations}
\indent Our analytic theory does not take into account the specific character of the atmospheric circulation, instead implicitly assuming that the circulation is dominated by day-to-night flow that acts to erase day-to-night pressure gradients. 
It should be utilized with caution in the limit of weak drag and large incident stellar flux, as it likely over-predicts the vertical diffusivity by an order of magnitude in this regime. Regardless, given that there is no analytic theory for what determines the strength of the superrotating jet itself, this theory is the sole analytic prediction for how vertical mixing rates should vary throughout parameter space. Additionally, the theory presented in this work applies only in the regime of chemical species whose chemical equilibrium abundance is the same on the dayside and nightside. When the background tracer abundance is not uniform, for instance in the case of a photochemically produced tracer on the dayside of the planet, the mixing does not act in a purely diffusive manner. As a result, this theory is not directly applicable to photochemical products in hot Jupiter atmospheres. \\
\indent Another example of a potentially non-diffusive tracer is cloud particles, which have a qualitatively different behavior compared to the chemical relaxation tracers used in this theory. Our theory is only for the chemical relaxation scheme, and we lack a theory for the aerosol case. Cloud models predict that clouds form preferentially in colder regions (for instance on the western limb and nightside of hot Jupiters), and hence the background distribution of cloud tracers is non-uniform. As shown by \cite{Zhang:2017}, assuming diffusion for a tracer with a non-uniform background mixing ratio can lead to an un-physically negative $K_\mathrm{zz}$. It is not yet clear if the vertical transport of cloud particles behaves diffusively, as it does for the chemical relaxation scheme considered in our theory. For the case of a tracer with a non-uniform background mixing ratio one needs to consider further non-diffusive effects to predict $K_\mathrm{zz}$ from first principles. As a result, we do not apply our analytic theory to predict the vertical mixing rates of aerosol tracers. Developing a theory for how aerosol settling in hot Jupiter atmospheres behaves (i.e., diffusively or non-diffusively) that predicts $\Kzz$ and how it would vary with particle size is an area for future research. 
\subsection{Application to observations}
We have developed a theory that is useful for understanding how vertical mixing rates vary over wide swaths of parameter space and shown that it agrees with numerical simulations of vertical tracer transport over a similarly large parameter space. To apply our theory to individual planets, the key unknowns are the tracer chemical lifetime, atmospheric composition, and frictional drag strength (if any). 
Using the theory presented here, one can estimate wind speeds and $\Kzz$ over orders of magnitude in parameter space. Our analytic theory can be used to predict $\Kzz$ for an individual planet, given a reasonable assumption for the values of $\tau_\mathrm{drag}$ and $\tau_\mathrm{chem}$. \\
\indent In this work, we have shown that the assumed frictional drag timescale can greatly affect atmospheric vertical mixing rates. Specifically, we found that including non-negligible frictional drag with a characteristic timescale of $10^4-10^6 \ \mathrm{s}$ can lead to an \textit{increase} in the vertical mixing rates for hot planets with equilibrium temperatures in excess of $1500 \ \mathrm{K}$. This is the first such study of the effect of frictional drag on vertical transport, as nominally it is assumed in GCMs of spin-synchronized gas giants which include passive tracers that there is no macroscopic drag in the observable regions of the atmosphere. As a result, we emphasize that future investigations of vertical transport in hot Jupiter atmospheres should consider this potential effect, as it can change their calculated $\Kzz$ by two or more orders of magnitude. Here drag represents additional forces, for example the Lorentz force associated with partial ionization in the atmospheres of ultra hot Jupiters or large-scale shear instabilities. Our work suggests that vertical mixing rates could be high in the situation where drag is relatively strong, but not so strong that it greatly reduces the strength of the atmospheric circulation. \\
\indent Recently, \cite{Sing:2015a} showed that cloud coverage varies greatly over an array of hot Jupiters with different planetary parameters. 
In general, we find that drag greatly affects vertical transport, which could lead to a large diversity in cloud coverage over the sample of hot Jupiters, as there are multiple possible drag mechanisms (e.g., Lorentz forces and shear instabilities) at play. Even for two planets with the same equilibrium temperature and rotation period, if they have different drag strengths (due to, for example, varied atmospheric compositions or internal field strengths) the vertical mixing rates in their atmospheres can differ by orders of magnitude. \\
\indent We also find from our numerical GCM simulations and theory that vertical transport is strongly dependent on the incident stellar flux, the aerosol particle size (for our aerosol tracers), and the chemical relaxation timescale (for our chemical relaxation tracers). For cool hot Jupiters with an incident stellar flux similar to HD 189733b, we find that aerosols of size $\lesssim 1 \ \mu\mathrm{m}$ can be lofted to $\sim$ mbar pressures. The lofting of aerosols to low pressures could explain the lack of molecular features in the near-IR observed for HD 189733b \citep{gibson2011,Gibson:2012aa,pont2013,Sing:2015a}, and our results are broadly in agreement with the particle size suggested from Rayleigh scattering at visible wavelengths \citep{Etangs2008}. \\
\indent In general, we find that planets that receive greater amounts of incident stellar radiation can more efficiently mix both chemical relaxation and aerosol tracers vertically. This is especially true for aerosol tracers, for which we find equatorial depletion in our GCM experiments with $T_\mathrm{eq} \le 2000~\mathrm{K}$. However, in our hotter simulations with $T_\mathrm{eq} \ge 2500~\mathrm{K}$, aerosol tracers are well-mixed latitudinally and longitudinally. We also find that the mixing ratios of chemical relaxation tracers at low pressures increases with increasing incident stellar flux. As a result, given a fixed chemical timescale, the relative abundance of species out of chemical equilibrium (e.g., CO/CH$_4$) would increase with increasing incident stellar flux. However, because chemical timescales decrease strongly with temperature, the atmospheres of ultra-hot Jupiters should be near chemical equilibrium \citep{Kitzmann:2018aa,Lothringer:2018aa}. \\
\indent Though in our suite of numerical GCM experiments we considered a broad range of incident stellar flux, frictional drag strength, aerosol tracer particle size, and chemical tracer relaxation timescales, future work could improve our understanding of vertical transport across planetary parameter space. The GCM utilized in this work has a simplified double-grey radiative transfer scheme, so studying vertical transport with a GCM that has more realistic radiative transfer would provide an improved understanding of vertical transport. Additionally, in this work we did not consider varying gravity, which would affect the mixing of aerosol tracers that would settle faster in the atmospheres of higher-gravity planets. We did not include the dependence of chemical relaxation timescales on temperature, which would cause the atmospheres of hotter planets to be closer to chemical equilibrium.  
Lastly, the analytic theory for vertical velocity and $\Kzz$ derived in this work only applies to chemical relaxation tracers. Future work could determine the non-diffusive effects arising from aerosol tracers and incorporate them into a theory for the vertical transport of aerosols in hot Jupiter atmospheres.
\section{Conclusions}
\label{sec:conclusions}
\indent In this work, we conducted a large suite of GCM experiments of hot Jupiter atmospheres including passive tracers to model chemical relaxation and the settling of aerosols. We studied how the mixing of these passive tracers depends on the incident stellar flux, drag strength, chemical timescale, and particle size. To understand these experiments, we developed analytic theory that predicts vertical wind speeds and vertical mixing rates as a function of planetary parameters. We compared this theory to our GCM experiments, finding good agreement over a wide range of parameter space. From this work, we draw the following key conclusions:
\begin{enumerate}
\item Vertical transport in hot Jupiter atmospheres is characterized by three distinct regimes, divided by the amount of frictional drag on the atmospheric flow. When there is little frictional drag ($\tau_\mathrm{drag} \gtrsim 10^7~\mathrm{s}$) and hence a strong superrotating jet, the vertical velocities have a columnar structure, with strong vertical transport occurring primarily in local regions where the vertical velocities are high. When frictional drag is moderate (with $10^6~\mathrm{s} \gtrsim \tau_\mathrm{drag} \gtrsim 10^5~\mathrm{s}$), the circulation is characterized by day-to-night flow, with upwelling everywhere on the dayside and downwelling on the nightside. When frictional drag is strong (with $\tau_\mathrm{drag} \lesssim 10^4~\mathrm{s}$), this day-to-night circulation is greatly damped due to the drag. We find that vertical transport is strongest in the moderate-drag regime, as there is a coherence between vertical velocity and tracer abundance, and yet the vertical velocity amplitudes are still relatively large -- in effect, a ``happy medium'' in which particle mixing is at its greatest. 
\item For both chemical relaxation and aerosol tracers, we find that tracer mixing ratios aloft increase with increasing equilibrium temperature (or equivalently, incident stellar flux). This is because vertical wind speeds increase with increasing equilibrium temperature, leading to stronger vertical transport of passive tracers. For our chemical relaxation tracers, we find that the tracer mixing ratio aloft increases with increasing chemical timescale. The increased tracer mixing with increased chemical timescale occurs because long chemical relaxation timescales make it harder for the chemistry to relax back to equilibrium when vertical motions transport chemical species from below; therefore, chemical disequilibrium (i.e., strong deviations from chemical equilibrium) are much greater when the chemical timescales are longer. For our aerosol tracers, we find that the tracer mixing ratio aloft increases with decreasing particle size. This is because smaller particles have a lower terminal velocity, and as a result are more easily vertically lofted by winds.  
\item We calculate the vertical diffusivity $\Kzz$ from our simulations with both chemical relaxation and aerosol tracers, finding that $\Kzz$ increases with increasing equilibrium temperature. For chemical relaxation tracers, $\Kzz$ generally increases with increasing chemical timescale, but there are deviations to this trend when the chemical timescale is comparable to or shorter than the dynamical timescale. For aerosol tracers, we find that $\Kzz$ is relatively independent of particle size, similar to the results of \cite{parmentier_2013}. The insensitivity of $\Kzz$ on particle size is due to the interaction of dynamics with aerosol mixing causing the ratio of the tracer flux and vertical tracer gradient to be approximately constant with changing particle size.  
\item From both analytic theory and numerical simulations, we find that vertical velocities increase with increasing equilibrium temperature and increasing atmospheric frictional drag timescale (until the drag timescale is longer than the rotation period of the planet, at which point the effect of drag weakens). Additionally, the theory of \cite{Komacek:2015} and \cite{Zhang:2016} matches quantitatively to a factor of $\sim 2$ the vertical velocities predicted from our suite of double-grey GCM experiments over more than four orders of magnitude in parameter space. 
\item Modifying the Earth-based theory of \cite{Holton:1986} to apply in spin-synchronized gas giant atmospheres, we used our analytically predicted vertical velocities to derive a first-principles, analytic prediction for the effective vertical diffusivity $\Kzz$. We find broad agreement between the $\Kzz$ estimated by our theory and that calculated in a suite of GCM experiments, though the theory over-predicts $\Kzz$ by an order of magnitude at high equilibrium temperatures ($\ge 2000 \ \mathrm{K}$) when frictional drag is weak ($\tau_\mathrm{drag} \ge 10^7 \ \mathrm{s}$). Given that we analytically predict the scaling of $\Kzz$ with equilibrium temperature and drag strength, this theory can be used to estimate the dependence of $\Kzz$ on planetary parameters for one-dimensional models of spin-synchronized gas giant atmospheres. We describe how to estimate $\Kzz$ using our analytic theory in \Sec{sec:cook}.
\item Our results have implications for current and future observations of aerosols and disequilibrium chemistry in hot Jupiter atmospheres. We find that the vertical transport of aerosol settling tracers is much stronger for smaller particle sizes. For cooler hot Jupiters with $T_\mathrm{eq} \lesssim 2000 \ \mathrm{K}$, aerosols $\lesssim 1 \ \mu\mathrm{m}$ in size can be mixed to mbar pressures. As a result, mixing of aerosols with small particle sizes provides an explanation for the absence of molecular absorption features in many hot Jupiter atmospheres. We also find that the vertical transport of chemical relaxation tracers is strongly dependent on their chemical relaxation timescale, incident stellar flux, and frictional drag strength. 
Future observations of hot Jupiter atmospheres with \textit{JWST} can probe the transition from cooler atmospheres with CO and CH$_4$ in chemical disequilibrium to hotter planets that have short chemical timescales relative to the dynamical timescale and as a result will be near chemical equilibrium.
\end{enumerate}
\acknowledgements
We thank Xianyu Tan and Xi Zhang for helpful comments on an early draft of this manuscript, and the anonymous referee for helpful comments that significantly improved the manuscript. 
T.D.K. and A.P.S. acknowledge support from the Technology and Research Initiative Fund (TRIF) at the University of Arizona. T.D.K. acknowledges funding from the 51 Pegasi b Fellowship in Planetary Astronomy sponsored by the Heising-Simons Foundation. 
\if\bibinc n
\bibliography{References_hj}
\fi

\if\bibinc y

\fi

\end{document}